\documentclass[16pt,a4]{article}
\usepackage{epsfig}
\usepackage{setspace}

\title{Size and Defect related Broadening of Photoluminescence Spectra in ZnO:Si Nanocomposite Films}

\author{Shabnam$^a$, Chhaya Ravi Kant$^a$ and P.
Arun$^b$\footnote{email:arunp92@physics.du.ac.in, Telephone:091 011
29258401, Fax: 091 011 27666220} \\ \\
$^a$Department of Applied Sciences,\\ 
Indira Gandhi Institute of Technology,\\ 
Guru Gobind Singh Indraprastha University,\\ Delhi 110 006, India.\\
\\
$^b$Department of Physics \& Electronics,\\ S.G.T.B. Khalsa College,\\
University of Delhi, Delhi - 110 007, India\\
}

\begin{document}
\maketitle

\begin{abstract}
Nanocomposite films of Zinc Oxide and Silicon were grown by thermal evaporation technique using varying ratios of ZnO:Si in the starting material. Analysis reveal the role of ZnO and amorphous silicon interface in contributing to relatively less common blue photoluminescence emissions (at $\sim$ 400 and 470nm). These blue peaks are observed along with the emissions resulting from band edge transition (370nm) and those related to defects (522nm) of ZnO. Careful analysis shows that along with the grain size of ZnO, a suitable compositional ratio (of ZnO to silicon) is critical for the coexistence of all the four peaks. Proper selection of conditions can give comparable photoluminescence peak intensities leading to broad-band emission. 
\end{abstract}

\vfil \eject

\section{Introduction}

Recent research in Material Science is directed along the lines of tailoring 
material properties as per requirement. In this direction both materials in 
nano regime and in composite state has drawn attention. Interest hence has 
also been developed in materials in nanocomposite state. Nanocomposites by 
defintion refer to the class of composite materials wherein atleast one of 
the constituents exists in the nanometre range. Research on ZnO based 
nanocomposites has been a front-runner in recent times, especially in 
the field of white light emitting devices (LED) \cite {znoEg, znoEg2}. 
These nanocomposites enjoy two-fold advantage over ZnO nanomaterials. While 
a carefully selected host material can result in broadening of light emission, 
it also gives additional stabilty to the ZnO film and prevents agglomeration 
of the grains. Of all the possible hosts, the research is mainly centred 
around the use of silicon and its varied complexes \cite{siuse1, siuse2, 
siuse3, siuse4, siuse6, siuse7}. This is because of the key position of 
silicon in the microelectonic industry and its abiliy to emit in the red 
region in the nanophase. 

\par In quest for obtaining white light, Klason et al \cite{kal} have tried 
depositing n-type ZnO nanorods on p-Si, which emitted white light under 
forward biasing. Depositing of ZnO nanoparticles in porous silicon (po-Si) 
has  also been reported by Bo et al \cite{z.bo} and Mehra et al \cite{mehra}. 
Though the intensity of luminescence observed in Bo's work was high, use of 
porous Si substrate itself suffers from the demerit of being prone to quick 
oxidation and requires tedious method of production. Peng \cite{ypeng1} and 
Pal \cite{pal}had studied the effect of varying ZnO to silicon content. Peng 
et al \cite{ypeng1} were successful in obtaining white light from RF 
sputtered films with the lowest ZnO content (26\% of ZnO). By considering the 
role of heterogeneous boudaries and taking into account the depletion width 
thus developed, they explained the maximum obtained intensity of the blue 
photoluminescence (PL) peak observed in the samples with lowest Zinc Oxide 
nanoparticles. In a later work \cite{ypeng2} they also discussed the color 
chromaticity dependance on the dot diameter. With their work they were able 
to prove that the neighbourhood and the size of the ZnO grains are both 
important for the efficient white light production. Using a similar method 
of fabrication Pal et al tried dispersing Si atoms in ZnO matrix. The 
samples were then annealed at high temperature. An increase in PL intensity 
was observed upon annealing and with decrease of ZnO content in the films. 
Unlike the PL behavior reported by Peng, the intensity in Pal's samples 
decreased  with further lowering of ZnO content of the films. Thus it 
becomes clear that the compositional ratio of ZnO to Silicon which inturn 
means the neighbourhood offered to ZnO nanoparticles is deterimental in 
bringing desired changes in optical properties of the nanocomposites and 
should be given its due concern. Nanocomposites based on ZnO and Si have 
been fabricated by sol gel \cite {solgel1, solgel2, solgel3, solgel4, 
solgel5}, RF sputtering \cite{sput1, sput2, sput3, sput4}, pulsed laser 
deposition \cite{pld1}, via chemical routes \cite{chemroute1, chemroute2, 
chemroute3, chemroute4}, reverse micelle method \cite{bouvy}and their 
optical properties studied. These research calls for more systematic work 
to optimize the parameteres for tailoring the properties and for a better 
understanding of interface and defect mechanism of ZnO:Si nanocomposites

\par Hence to this effect, we have started systematic study of Zinc Oxide 
grains in nano regime embedded in amorphous Silicon matrix. In our earlier 
work,\cite{paper1} we had demonstrated that these films emitted light in 
the visible region, namely in UV, blue, green and red regions. To investigate 
the tailoring of our samples' properties, we annealed our samples (post 
deposition) and reported broadening of photoluminescence emissions. Not 
only this, the intensity of the annealed samples increased manifold 
\cite{paper2}. In this manuscript we report the role of varying ratio 
of ZnO:Si (in starting material) on the photoluminescence.

\section{Experimental Details}

\par The films of ZnO:Si nanocomposites studied in this work were fabricated 
by thermally evaporating a mixture of powdered ZnO and n-Silicon. Vacuum of 
the order of ${\rm 10^{-6}}$ Torr was created in the deposition chamber of a 
Hind High Vac (Bengaluru), Thermal evaporation coating unit, Model 12A4D. 
The deposition was carried out on microcopic glass substrates maintained at 
room temeprature. To prevent the flying off, of starting material powdered 
ZnO and Si, was pelletized by applying a pressure of 60kN. ZnO used in this 
study was(99.99~\%) pure and purchased from Merck (Mumbai). The mixture was 
prepared by mixing ZnO and Silicon in the proportions of 1:1, 1:2, 1:3, 2:3 
and 2:5 (by weight). Films of these composition are hereafter referred to as 
sample (a), (b), (c), (d) and (e), respectively. We have restricted the 
present study to samples of 600\AA\, thickness. The samples discussed in 
this work are as grown, not subjected to any post deposition heat treatment.

\par The structural studies of the surface is measured by Pananalytical 
PW3050/60 Grazing Incidence angle X-Ray Diffractometer (GIXD) and that of the 
bulk region by Philips PW 3020 X-Ray Diffractometer (XRD). X-Ray 
Photoelectron Spectroscopy (XPS) was performed with Perkin-Elmer X-ray 
Photo-electron Spectrometer (Model 1257) with Mg ${\rm K\alpha}$ (1254 eV) 
X-ray source. Photoluminescence (PL) scans were recored on Jobin Yvon 
spectroscopes respectively. Renishaw's ``Invia Reflex'' Raman spectroscope 
was used for measurements. The surface morphology and texture of the as 
grown nanocomposite films were studied using the Scanning Electron 
Microscope (SEM), atomic force microscopy (AFM)- NTEGRA NS-150 
and Transmission Electron Microscope (TEM). In the following section we 
enlist the results of the various analysis done on our samples.

\section{Results and Discussions}

\subsection{X-Ray Diffraction, Chemical Composition \& Morphological Studies}
The nanocomposite films were fabricated by co-evaporating Zinc Oxide  with 
Silicon. The melting point of silicon is ${\rm 1410^oC}$ at atmospheric 
pressure while that of Zinc Oxide is about ${\rm 1975^oC}$. However, the 
melting point of Zinc Oxide is known to reduce in the presence of group-IV 
elements facilitating its thermal evaporation at a lower temperature 
\cite{lowmp1,lowmp2,lowmp3}. Both these materials have diverse melting 
points which could result in multilayered structure rather than a 
homogeneous nanocomposite film. The possibility of formation of multilayered 
structure was overruled by IR analysis of the films which was reported in 
our earlier work \cite{paper1}.

\par We have further tried to resolve this question using grazing incidence 
angle X-Ray diffraction studies (GIXD). At critical incidence angle, usually 
lying between ${\rm 1.5^o}$ and ${\rm 0.005^o}$, the X-ray beam gets totally 
reflected from the surface. Infact below the critical angle, the reflected 
beam is allowed to penetrate only upto few nanometres \cite{gixd1, gixd2, 
gixd3, gixd4}. This ensures X-Ray is being diffracted from the surface layer. 
Figure 1A exhibits the nature of the diffraction pattern obtained for films 
of the same thickness but varying ratios of Zinc Oxide to Silicon. These 
X-Ray diffraction pattern show prominent peaks at ${\rm 2\theta=43^o}$, 
${\rm 36^o}$ and ${\rm 38^o}$. These peaks correspond to pure Zinc (ASTM 
Card No-4-831). The ${\rm 36^o}$ peak can possibly be unresolved peak of 
Zinc and Zinc Oxide (ASTM Card No-1451). The existence of elemental Zinc as 
shown by GIXD would imply that the surface of our films have metallic Zinc. 
This may not be the case within the surface of the films. To investigate 
this we have also studied the same samples using ${\rm \theta-2\theta}$ 
X-Ray diffraction studies.  The results are shown in figure 1B. Barring 
sample~(a) corresponding to the 1:1 ratio film, the X-Ray diffraction pattern 
of all the other samples were unmarked. This did not come as a surprise since 
the grain size of ZnO in nanocomposite films of ${\rm 600\AA}$ is too small 
to result in X-Ray diffraction \cite{paper1, gwo}.

\par In short, our X-Ray analysis suggests formation of pure Zinc on the 
surface of nanocomposite bulk, where Zinc Oxide is embedded within the matrix 
of Silcon. To investigate the depth to which elemental Zinc exist, we have 
analysed sample (a) using XPS. XPS scans were made at the surface and 
periodically repeated after removing layers of 100\AA \, by ion milling. 
Figure 2A shows the Zinc 1022eV peak associated with it ${\rm 2p_{3/2}}$ 
orbital. Figure 2A compares the Zinc peaks as scanned at depths of 100, 
200, 300, 400 and 500\AA. These are single peaks which show a shift to lower 
energy side with increasing depth. However, as can be seen from Figure 2B, 
the Zinc peak from the surface layer can be deconvoluted into two peaks. The 
peaks are separated by $\sim$1eV. The 1021eV is associated with free Zinc or 
Zinc atoms which are not in bonding with any other species \cite{xpswagner, 
briggs, xps1, xps2, xps3}, while the 1022eV peak shows that there are 
sizeable amount of Zinc atoms that present in bonding with oxygen, existing 
as Zinc Oxide \cite{xpswagner, briggs}. To colloborate this, we have also 
scanned the layers for oxygen.

\par  Figure 2C compares the oxygen peaks along the thickness of the film. 
Here also, oxygen seems to exist in two chemical compositional states as is 
indicated by the deconvoluted peaks shown in the figure. The only possibility 
is oxygen in bonding with Zinc and Silicon present in the film. The 530.5eV 
peak of the surface layer, corresponds to oxygen in bonding with Zinc, 
confirming that the surface of sample (a) has Zinc and Zinc Oxide. A close 
look of Figure 2C, reveals that the oxidation of silicon at the surface has 
not passivated the surface and protected the inner layers completely, as can 
be understood by the presence, abid diminishing contribution of the 532eV 
peak. This could be because the sample in question is of small thickness and 
the XPS scanning was done after a long time from its date of fabrication. 
Figure 2D exhibiting Si XPS peaks, also confirms the existence of Silicon 
and Silicon dioxide in the film with the deconvoluted peaks clearly 
observable through the entire thickness of the sample (a). This establishes 
that our film contain Zinc Oxide embedded in a matrix of Silicon and Silicon 
Dioxide with elemental Zinc at the surface. However, it must be emphasised 
again that the occurrence of silicon dioxide only took place after a long 
time from the sample's fabrication time. 

Before proceeding it is worth mentioning again that the Zinc XPS peak 
corresponding to atoms in bonding with oxygen shifts with their position 
along the depth of the film. Figure 3 shows the near linear trend with this 
peak moving from 1022.4eV at the surface to 1021.8eV within the film. We 
beleive this shift in peak as you go within the thickness of the film is due 
to the chemical species and it's relative quantity existing around the ZnO 
molecules. Figure 3B clearly shows increase in relative Zinc Oxide as you 
move deeper into the film. It should be noted that our starting material for 
sample (a) had 50\% ZnO however as can be anticipated, the film's bulk only 
has 24-30\% ZnO. Figure 3C combined the best fit curves of Figure 3A and 3B 
and projects variation in peak position as ZnO percentage varies. Larger 
amount of Silicon seems to shift the ZnO peak to the higher energy side. 
This is the precise influence we are investigating and hope would assist 
in tailoring the optical properties of these nanocomposites. Also, we expect 
the same  trend of Figure~3C for samples prepared with differnt ZnO presence. 
We are in process of studying samples with different starting ratios, same 
thickness and verifying this. 

\par TEM micrograph (Figure~4) of sample (c) shows existence of two phases. 
The clusters of ZnO dispersed in a matrix of amorphous Silicon. The inset 
shows image taken in dark field which brings better contrast. For studying 
the film-morphology we carried out AFM analyis on the sample surface. The AFM 
images of the samples (a) and (b) are shown in Figure 5A and 5B, respectively. 
The AFM images reveal uniform films with particles ranging in nanometre. 
However, while discussing the X-Ray analysis results, we have clearly 
established the existence of elemental Zinc on the surface of the films 
hence, the observed grains and measured grain size are not typically of 
ZnO. Hence, the grain size of ZnO was also calculated using the Debye-Scherrer 
formula \cite{cullity} from the ${\rm 36^o}$ peak of GIXD. The grain size of 
ZnO was found to lie between 8.8 to 23nm for our samples.

\par Along with the nature of Zinc Oxide's neighbouring chemical composition 
and it's grain size, it is understood that the strain acting on Zinc Oxide 
lattice also contributes to its optical properties. The ${\rm 36^o}$ ZnO 
peak in XRD is only observed in one sample (that of `a', see Figure 1B). 
There is a distinct right shift in the observed peak position with respect 
to the peak position reported for a single crystal. This shift is indicative 
of a tensile stress. We have calculated the strain from which stress can be 
calculated \cite{cullity}. Strain is usually evaluated using \cite{cullity} 
\begin{eqnarray}
\sigma={d_{ASTM}-d_{obs}\over d_{ASTM}}\label{eq1}
\end{eqnarray}
where ${\rm d_{ASTM}}$ is the `d' spacing for the single crystal structure 
and ${\rm d_{obs}}$ is that for the sample in question. Using equation (1) 
the strain within the bulk of sample (a) is found to be ${\rm 3.054 \times 
10^{-3}}$. As expected, this is different from strain acting on the surface, 
which was found to be  ${\rm 7.372 \times 10^{-3}}$. However, since the order 
is same we can consider strain acting on surface is representative of that in 
the bulk. The surface strain was found to be decreasing with decreasing ZnO 
presence in the films. The tensile strain detected is due to the defects 
caused by vacancies \cite{tensilevac}. This can also be appreciated from 
Raman spectra. Hence, we analysed our samples using Raman Spectra. The 
results of our study is discussed below.

\subsection{Raman Spectra}

The Raman spectra were taken in standard back scattering geometry using 
${\rm Ag^{2+}}$ laser for excitation. Figure 6 gives the Raman spectra 
for the sample (a), (d), (b), (e) and (c) respectively. Broad peaks were 
obtained from ${\rm 300cm^{-1}}$ to ${\rm 600cm^{-1}}$. These peaks were 
deconvoluted to give three peaks. We are confident about our deconvolution 
at ${\rm \sim 310cm^{-1}}$ and ${\rm \sim 570cm^{-1}}$ because of the 
prominent shoulders they give. The ${\rm \sim 310cm^{-1}}$ peak corresponds 
to the LA mode of amorphous Silicon \cite{paper1, a:siraman1}. The intensity 
and contribution of this peak to the Raman spectra is indicative of the 
amorphous Silicon in our samples. The large Silicon shown in Raman analysis 
which were carried out soon after the fabrication confirms our arguement that 
Silicon Oxide (shown in XPS) formed with time. 

The ${\rm \sim 570cm^{-1}}$ corresponds to the A1 LO peaks of ZnO which is 
essentially associated with defect structures in ZnO. This defect in 
literature has been associated to either Oxygen vacancies or Zinc 
interstitial defects \cite{zniraman}. From our X-Ray analysis on we reported 
vacancies giving rise to tensile strain. Hence, we beleive this ${\rm 
570cm^{-1}}$ peak is due to Oxygen vacancies in our samples. This is ofcourse 
expected to influence the optical properties of the samples. The third peak 
is at ${\rm \sim 430cm^{-1}}$. Although the position of this deconvoluted 
peak may not be as reliable as the other peaks because of the lack of 
prominent shoulders, this can be ascribed as the peak of ZnO. The peak at 
${\rm 430cm^{-1}}$ is of ${\rm E_2^{high}}$ mode of Zinc Oxide, usually 
reported at ${\rm 437cm^{-1}}$ \cite{raman1, raman2, raman3} and possibly 
merged with amorphous Silicon that is expected at ${\rm 470-510cm^{-1}}$ 
\cite{a:siraman1, a:siraman2, a:siraman3}. The ${\rm E_2^{high}}$ peak is 
related to the vibration of the oxygen atom in the wurtzite structured ZnO 
lattice. The presence of this huge peak in all the samples further confirms 
the presence of ZnO in the bulk region of the films. The existence of 
${\rm 430cm^{-1}}$ reconfirms our claim of ZnO being embedded in the 
amorphous Silicon.

It is clear from the Raman Spectra that our nano-composite samples is not 
just ZnO clusters dispersed in a matrix of amorphous Silicon but additionally 
the ZnO cluster is a homogeneous mix of lattice with perfect wurzite 
crystallinity and that with oxygen defects. Considering that the optical 
properties of ZnO strongly depends on it's structure giving PL emission at 
520 or 370nm depending on existence of defects or good crystallinity, 
existence of both phases in our sample should result in both peaks occuring 
in our PL. Thus, a suitable ratio of the contents of these two phases should 
give rise to a broad emission in the PL. Before looking at PL, we plot the 
ratio of area enclosed by the Raman peaks at ${\rm 430cm^{-1}}$ and 
${\rm 560cm^{-1}}$ with respect to the content of ZnO in the film (fig~7). 
Figure~7 shows that the two areas are comparable in samples (b), (d) and (e). 
Hence, we expect the PL emission peaks at 370 and 520nm to be comparable in 
these three samples. In our next section we shall investigate these issues.

\subsection{Photoluminescence}

\par To study how the above discussed strcutural and compositional behaviour 
manifests in the optical properties we have studied the PL spectroscopy of 
our samples using an excitation source of Xenon lamp. The PL was scanned in 
the range of 290-800nm on excitation with wavelength of 270nm. Since the 
measurements were taken up without the use of filters, the presence of second 
harmonics at 540nm is inevitable and can be seen as line spectra along with 
the PL peaks (Figure~8). The PL of samples (a), (b), (c) and (d) exhibit 
similar behavior with four prominent peaks. There is a sharp peak at 370nm 
accompanied by broad peaks at 410nm, 470nm and 522nm. The 370nm is due to 
the radiation resulting from electron transistion across the band gap of 
Zinc Oxide \cite{znoEg, znoEg2}. The 522nm is well documented 
\cite{greenzno1, greenzno2, greenzno3} and is related to intra-band 
transistions, i.e. between levels created by defects within the band gap, 
caused by the Oxygen vacancies. As discussed in the Raman analysis, we had 
predicted broadening of emission spectra in samples (b), (d) and (e) because 
they had nearly equal amount of ZnO lattice with and without defects. The 
expected broadening is evident in our PL (fig~8).

Broadening is also assisted in our PL due to additional peaks at 410 and 
470nm. The 414nm peak have been attributed to inter-grain boundary defects 
\cite{paper2} in ZnO. However, over the large amount of samples we have 
studied we have observed that the 410 and 470nm peaks appear only if the 
background is a-Si. These peaks were absent in samples with n-Si background 
\cite{paper1}. Hence, we propose that these peaks are indeed influenced by 
the intergrain boundary and the background species rather than, if not only 
by the defects. To further our idea, we argue that if these peaks are 
contributed by interface and hence by the net surface area of the ZnO 
grains, then the peak intensity would be a result of the grain size and 
number of grains present. This ofcourse is the underlying principle of 
nanotechnology. Assuming the number of grains in the film (N) to be 
proportional to the ratio of ZnO present in the starting material and 
using the grain size(R) evaluated using debye-Scherrer equation, we plotted 
a curve between the peak intensity of 465nm with ${\rm R^2N}$ (fig~9). We 
find this to be linear suggesting its dependence on the interface of ZnO and 
amorphous Silicon. 

\par Similar correlation is found among results of other analysis. For 
example, the 522nm PL peak arises due to oxygen vacancy defects. The defects 
also manifest as strain in the film as seen by the displacement of XRD peak 
from expected position. The perfect linear relation between the area of the 
green PL peak with strain in samples, (fig~9b) confirm our inferences. The 
omission of the 365nm peak also shows proportionality with amount of wurtzite 
species present as seen from Figure~9(c) which plots area under 365nm peak of 
PL with respect to ${\rm 438cm^{-1}}$ peak from Raman spectra. We can thus 
summarise that a sample with appropriate mixture of defects and defect free 
ZnO, grown with a starting material of 1:2 ZnO to Silicon with grains in nano 
regime would give four emission peaks that overlap. The overlapping results 
in broadening best understood by a schematic shown in Figure~10.

\section{Conclusion}
\par We have fabricated nanocomposite films of ZnO and Si using thermal evaporation technique with clusters of ZnO dispersed in a matrix of amorphous Silicon. Depth profiling ESCA scans along with earlier IR studies confirm formation of our nano-composite films. The PL of the samples shows four prominent peaks in the blue-green region, namely $\sim$ 400, 470, 370 and 522nm. The latter peaks are attributed to emissions via transistions from the band edge and defect related peaks from ZnO. On the other hand the 400 and 470nm peaks exist because of the interface between ZnO grains and amorphous silicon. The interface increases with decreasing grain size and increasing number of ZnO grains. This was established from the observation of varying PL intensity with varying the ZnO:Si compositional ratio and grain size. Samples with appropriate mixture of defects and defect free ZnO nanostuctures lead to broadening due to merging of peaks. We have found the 1:2 sample with maximum broadening. This sample with 22.4nm grain size also gave rise to large 465nm peak in PL indicating formation of large interfacial boundary. It thus proves, that not only the grain structure but also the interfacial boundaries and the nature of background species play a decisive role in equalizing the intensity accompanied by the broadening of the PL peaks. At present, we only observe a weak red emission and hope to control and improve the intensity of this peak. Enhancement and ultimately merging of red peak with the remaining emission spectra would estiblish ZnO:Si nano-composites as potential candidate for the development of cost effective white light emitting diodes.

\section*{Acknowledgment}
The resources utilized at Indian Institute of Technology, Delhi and Geology Department, University of Delhi is gratefully 
acknowledged. Also, the resources used at the Instrumentation Center and
University Information Resource Center, Guru Gobind Singh Indraprasta
University is also acknowledged. We also would like to express our gratitude 
to Dr. Kamal
Sanan and Dr. Mahesh Sharma (both at National Physical Lab., Delhi) for carrying out the photoluminescence and XPS studies respectively. Authors PA and CRK are thankful to University Grants Commission 
(India) for 
financial assistance in terms of Major Research Award, F.No-33-27/2007(SR).

\newpage
\section*{Figure Captions}
\begin{itemize}
\item[1.] X-Ray Diffractograms of sample (a),(b),(c),(d) and (e) obtained by 
(A) GIXD and (B) ${\rm \theta}$-${\rm 2\theta}$ technique.
\item[2.] (A) XPS ${\rm 2p_{3/2}}$ peaks of Zinc taken at various depth of 
same sample showing two species, (B) surface scan, (C) depth profile scans 
for Oxygen and (D) Silicon.
\item[3.] Variation of (A) Peak position of Zn with depth (B) Fraction of 
Zinc in bonding with Oxygen to amount of Silicon present along the thickness 
and (C) Peak position of Zinc in bonding with Oxygen to its fraction of 
presence.
\item[4.] TEM of sample (c). Inset shows dark field image of same sample taken 
at low magnification which clearly shows formation of two phase with ZnO 
clusters embedded in amorphous silicon.
\item[5.] AFM images of sample (a) and sample (b).
\item[6.] Raman spectra of sample (a), (d), (b), (e) and (c). Also seen are 
deconvoluted peaks assigned to amorphous silicon, wurtzite structure ZnO and 
with oxygen vacancies defects.
\item[7.] Relative presence of ZnO with oxygen vacancies to wurtzite 
structure ZnO (Area ${\rm 560cm^{-1}}$/Area ${\rm 438cm^{-1}}$ from Raman 
spectra) for varying ZnO content in film.
\item[8.] PL of sample (a), (d), (b), and 
(c). Alongside the raw spectra are shown, deconvolution give 365, 400 and 
465nm between 300 and 480nm. Also green emission due to defects have been 
separated from $\rm 2^{nd}$ harmonic to show relative contributions.
\item[9.] The (A) increase in contribution of 465nm PL peak with 
increasing ${\rm R^2N}$, (B) linear relation in green (522nm) emission with 
strain in film and inturn oxygen vacancies and (C) co-relation in existence 
of wurtzite peak (${\rm 438cm^{-1}}$ in Raman spectra) and blue emision 
(365nm peak of PL) in samples.
\item[10.] Schematic explaining broadening of PL in ZnO:Si nanocomposites and 
their individual contributions.
\end{itemize}

\vfil \eject

\begin{figure}[h]
\begin{center}
\epsfig{file=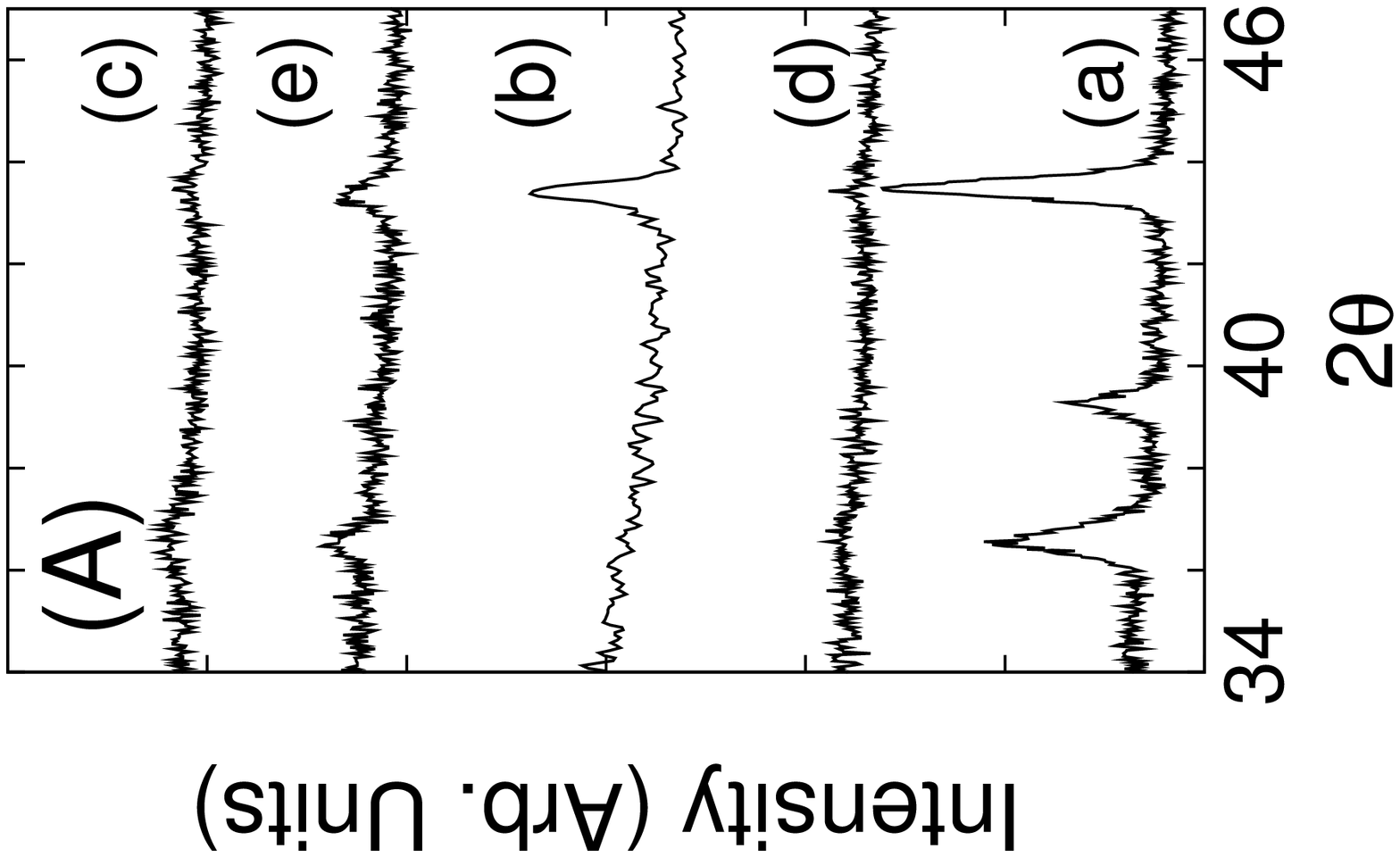, width=2.5in, angle=-90}
\hfil
\epsfig{file=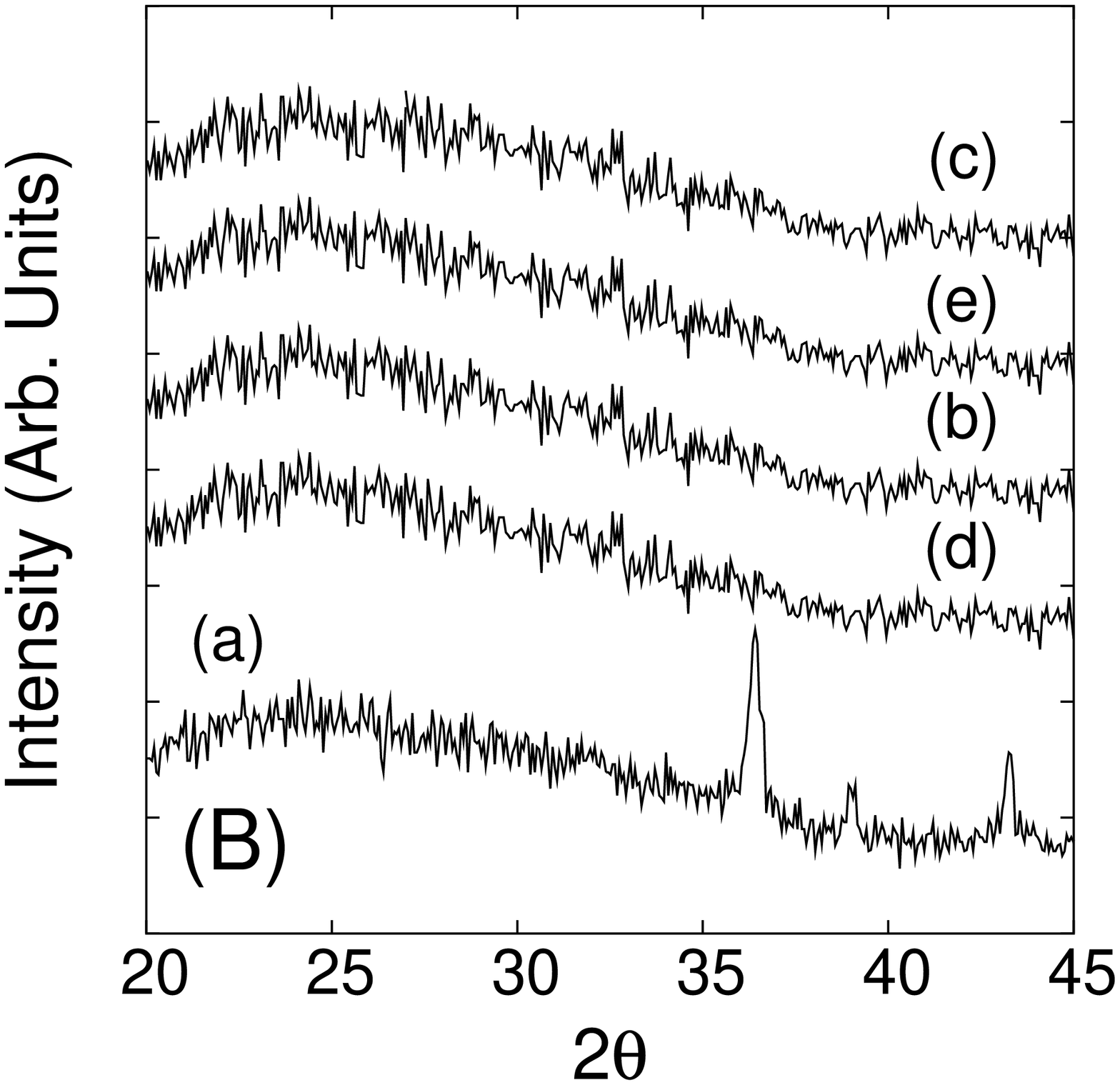, width=2.5in, angle=-90}
\end{center}
\caption{\sl X-Ray Diffractograms of sample (a),(b),(c),(d) and (e) obtained by 
(A) GIXD and (B) ${\rm \theta}$-${\rm 2\theta}$ technique.
}
\end{figure}

\begin{figure}[h]
\begin{center}
\epsfig{file=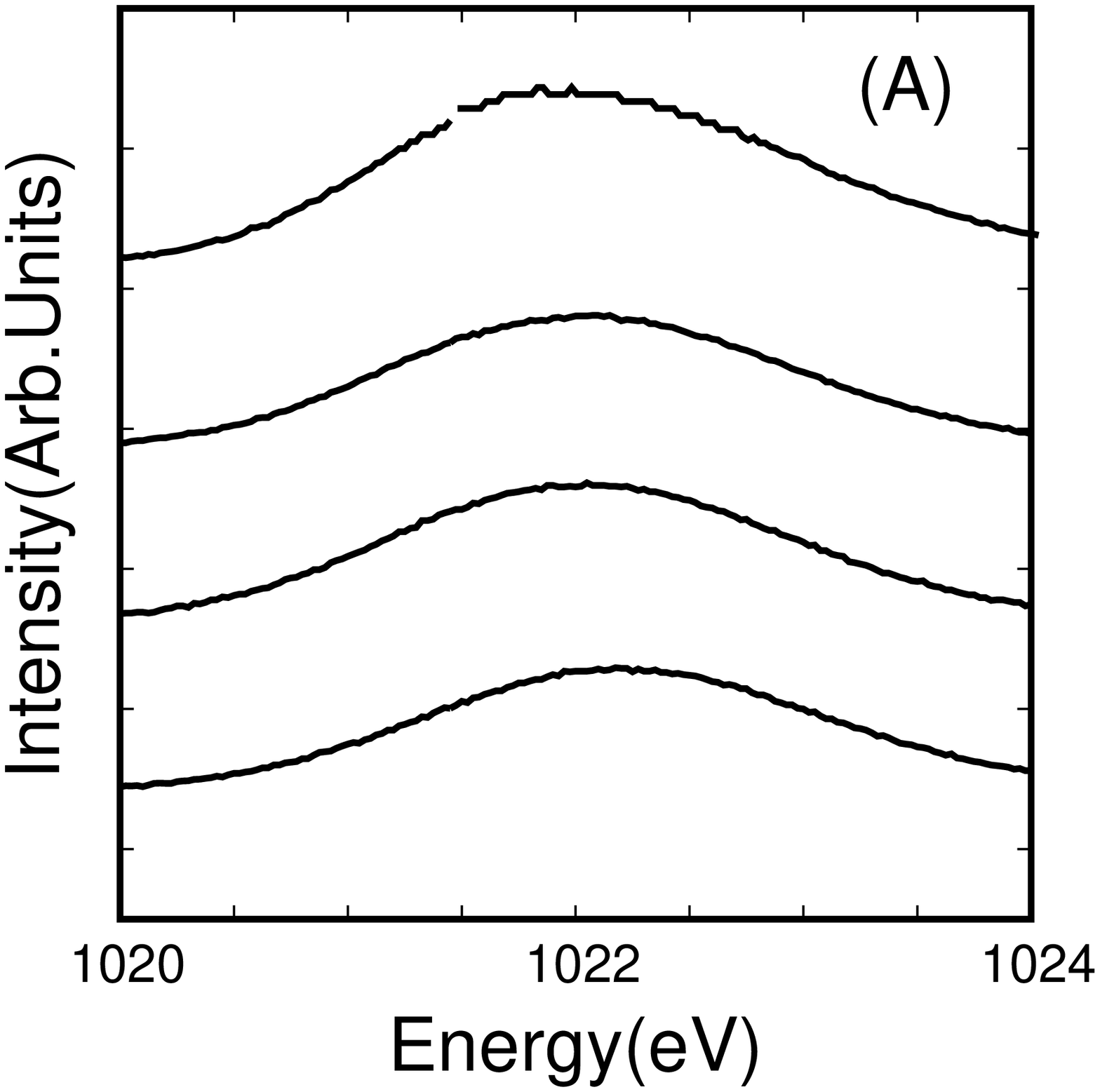, width=2.25in, angle=-90}
\hfil
\epsfig{file=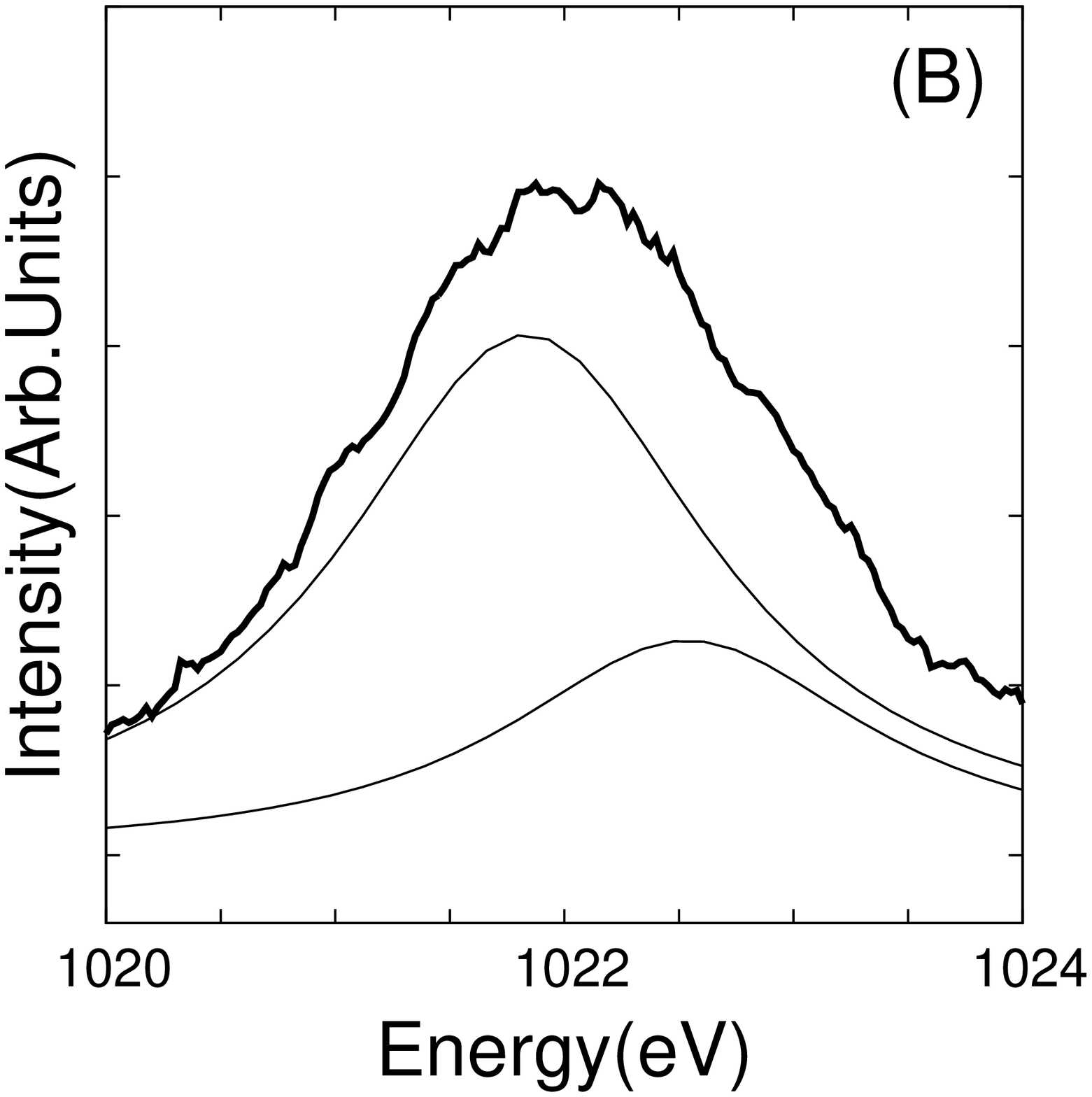, width=2.25in, angle=-90}
\vfil
\vskip 0.5cm
\epsfig{file=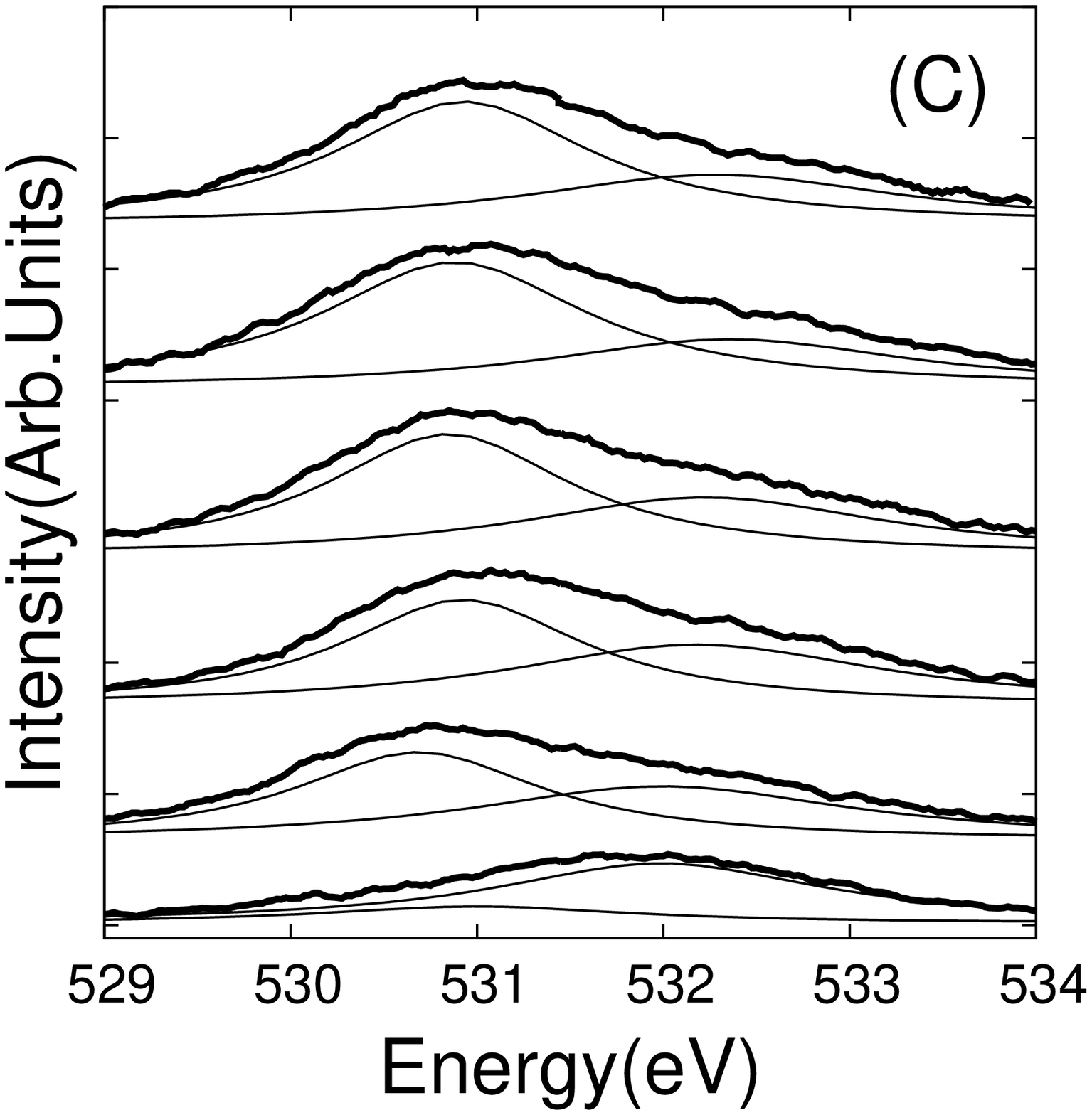, width=2.25in, angle=-90}
\hfil
\epsfig{file=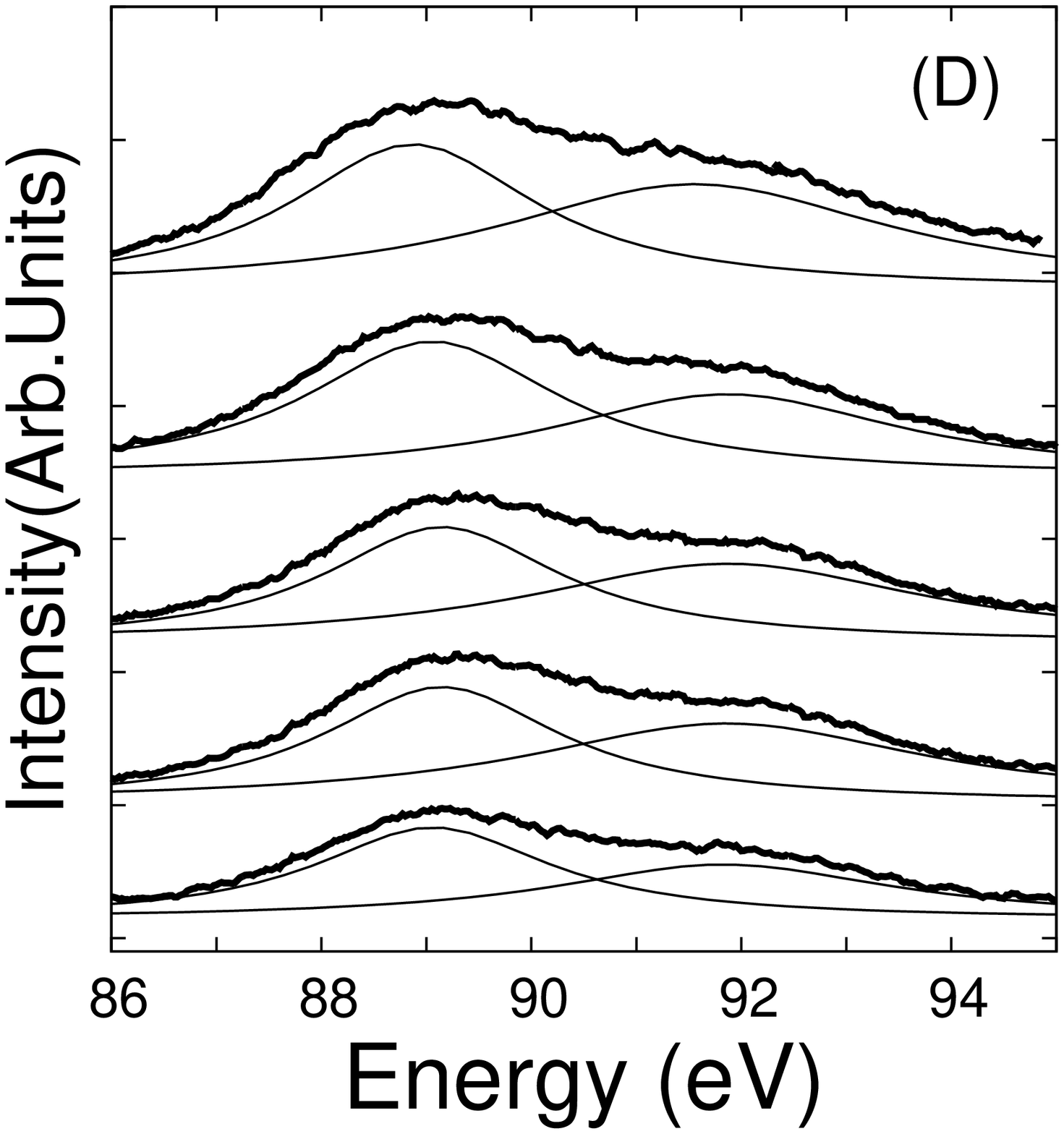, width=2.25in, angle=-90}
\end{center}
\caption{\sl (A) XPS ${\rm 2p_{3/2}}$ peaks of Zinc taken at various depth of 
same sample showing two species, (B) Surface scan, (C) Depth profile scans 
for Oxygen and (D) Silicon.}
\end{figure}

\begin{figure}[h!]
\begin{center}
\epsfig{file=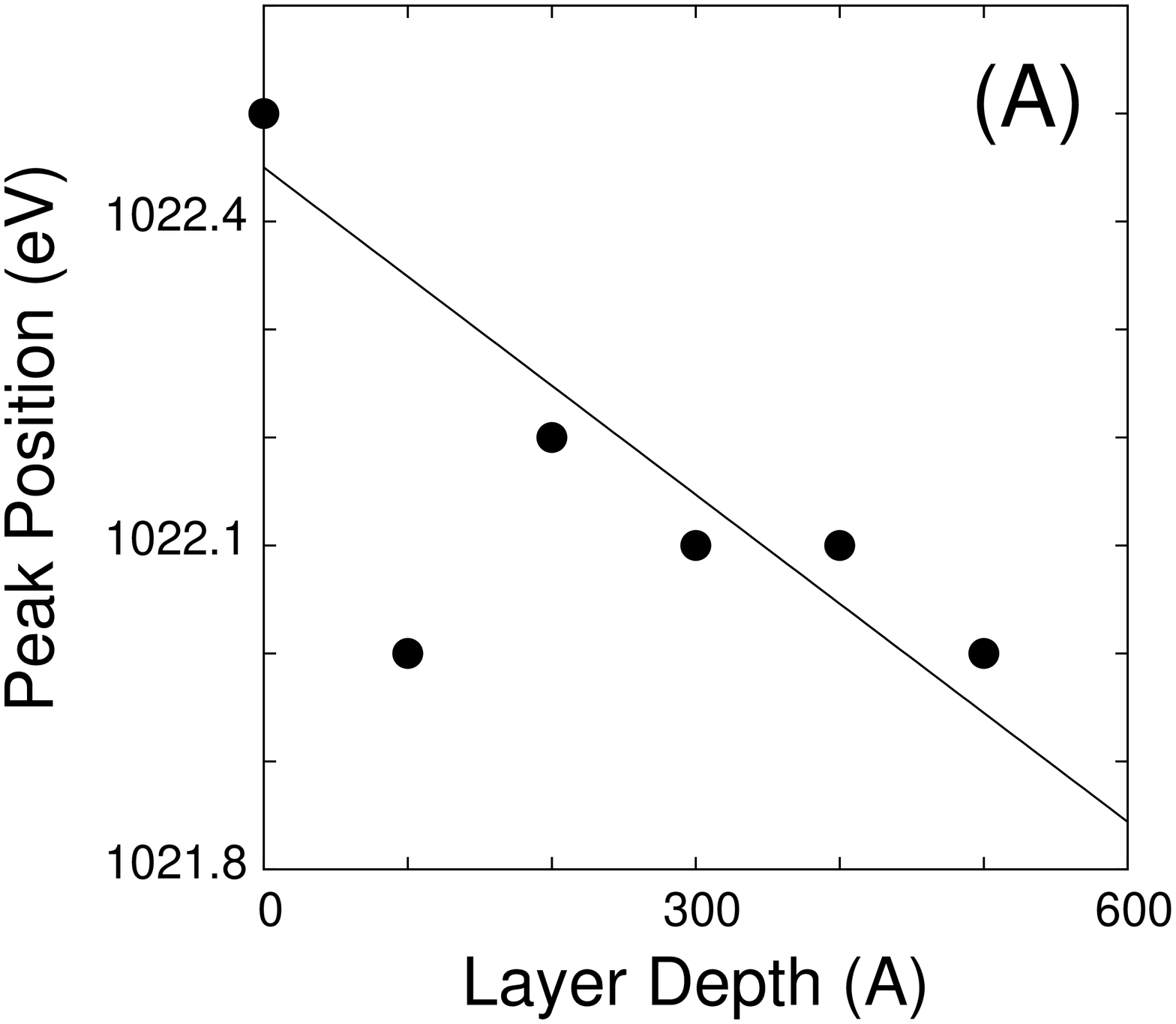, width=2.15in, angle=-90}
\vfil
\epsfig{file=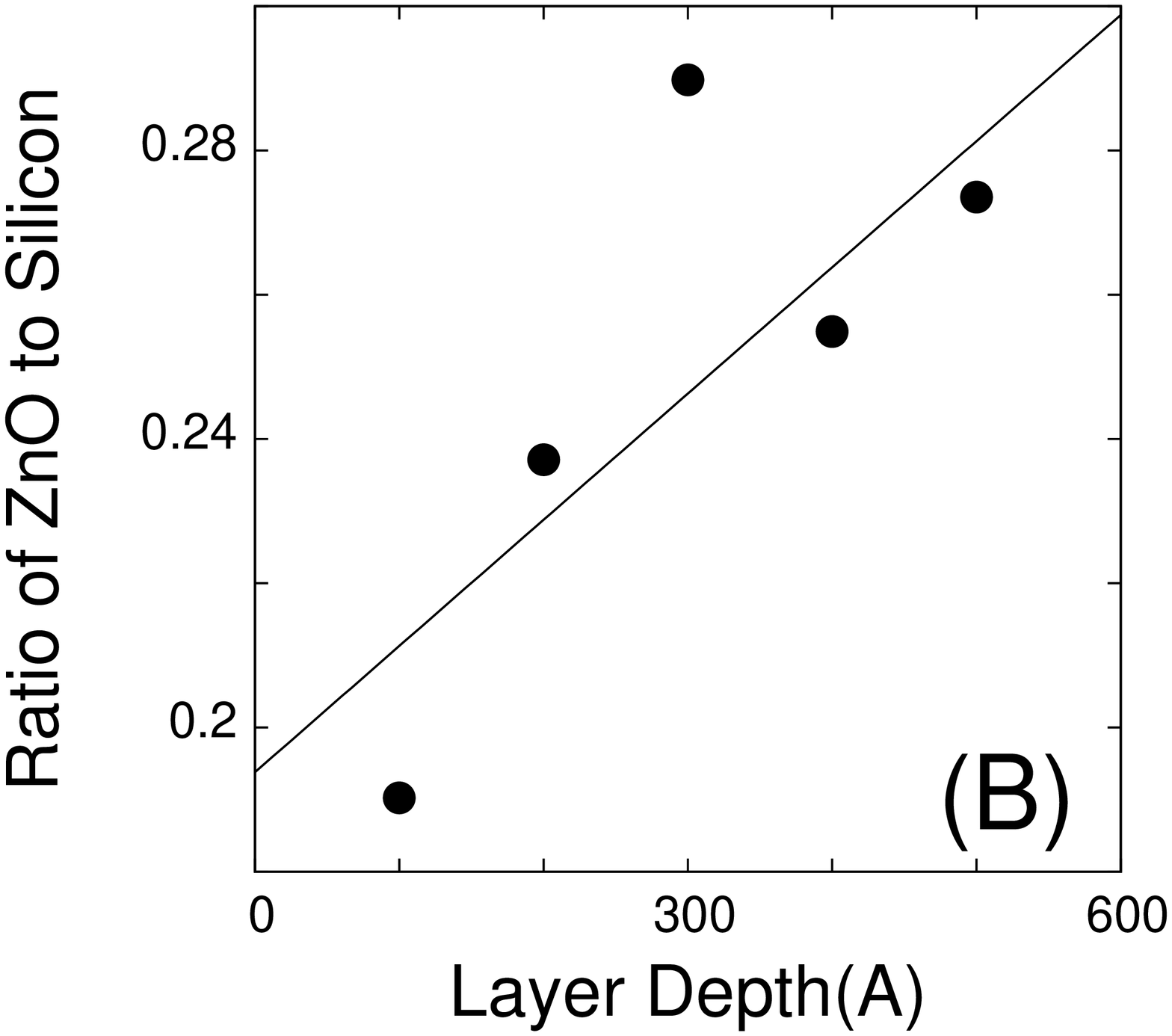, width=2.15in, angle=-90}
\vfil
\vskip 0.5cm
\epsfig{file=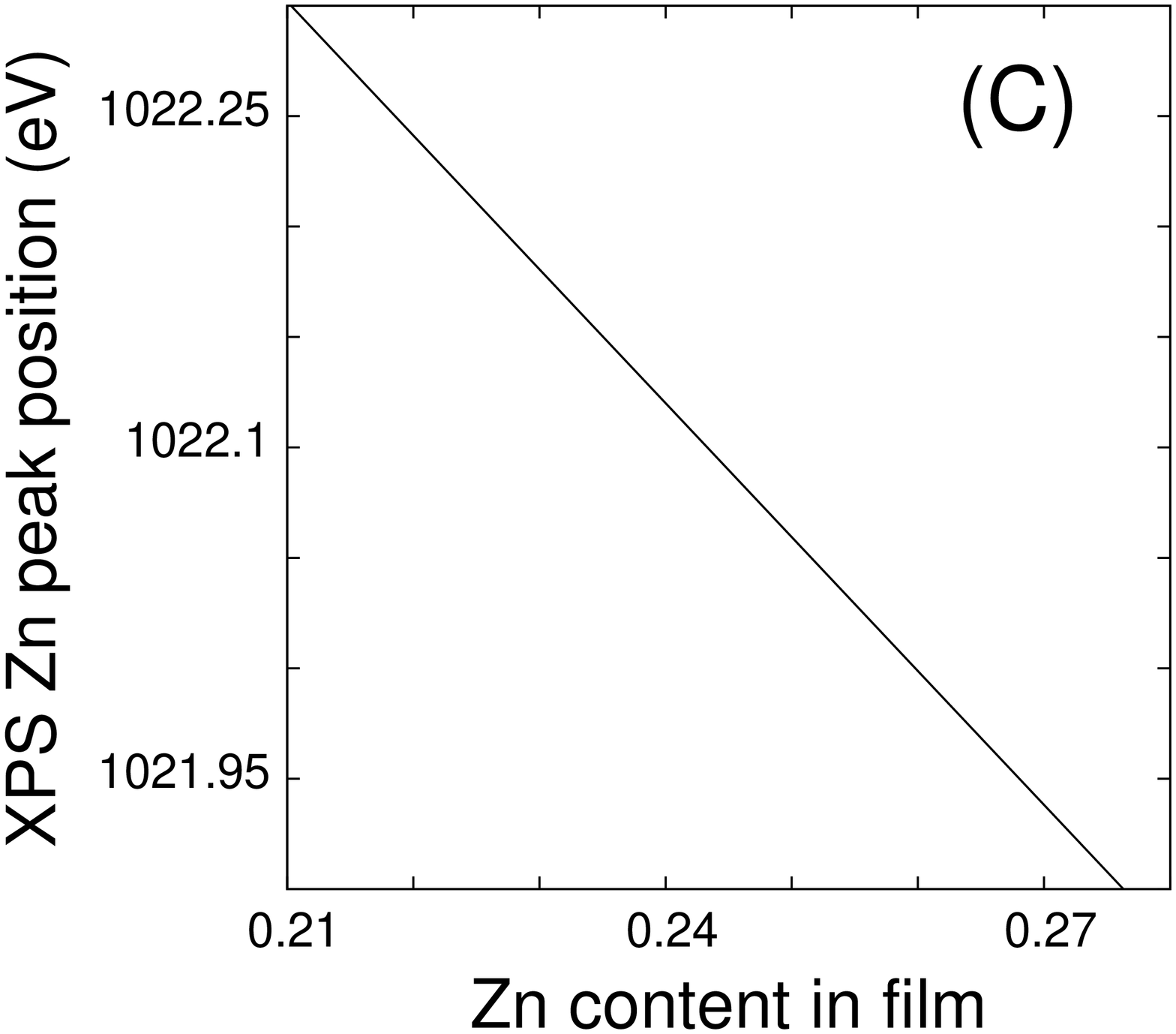, width=2.15in, angle=-90}
\end{center}
\caption{\sl Variation of (A) Peak position of Zn with depth (B) Fraction of 
Zinc in bonding with Oxygen to amount of Silicon present along the thickness 
and (C) Peak position of Zinc in bonding with Oxygen to its fraction of 
presence.}
\end{figure}

\begin{figure}[h]
\begin{center}
\epsfig{file=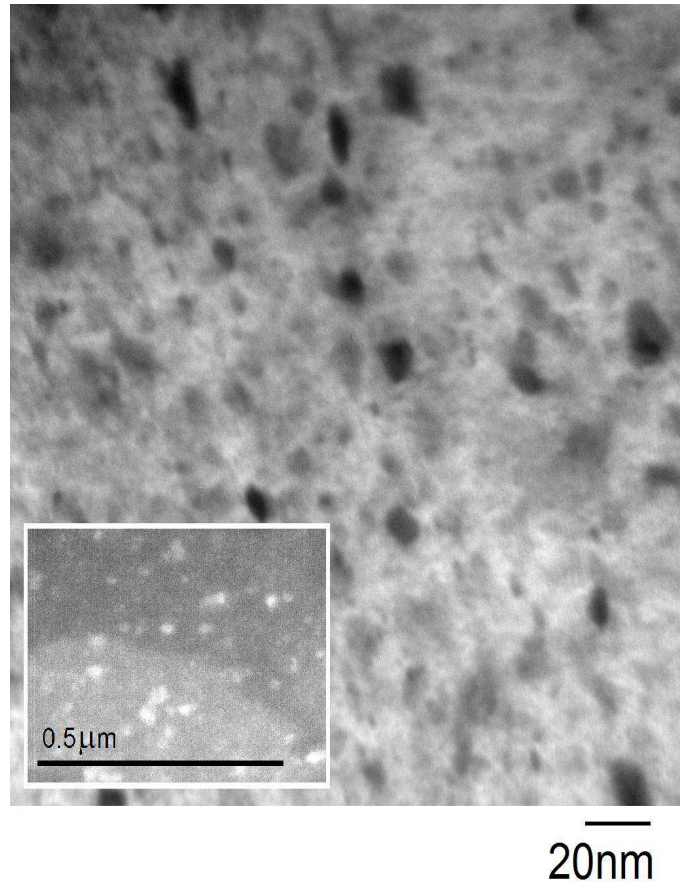, width=3.5in, angle=-0}
\end{center}
\caption{\sl TEM of sample (c). Inset shows dark field image of same sample 
taken at low magnification which clearly shows formation of two phase with ZnO 
clusters embedded in amorphous silicon.}
\end{figure}

\begin{figure}[h]
\begin{center}
\epsfig{file=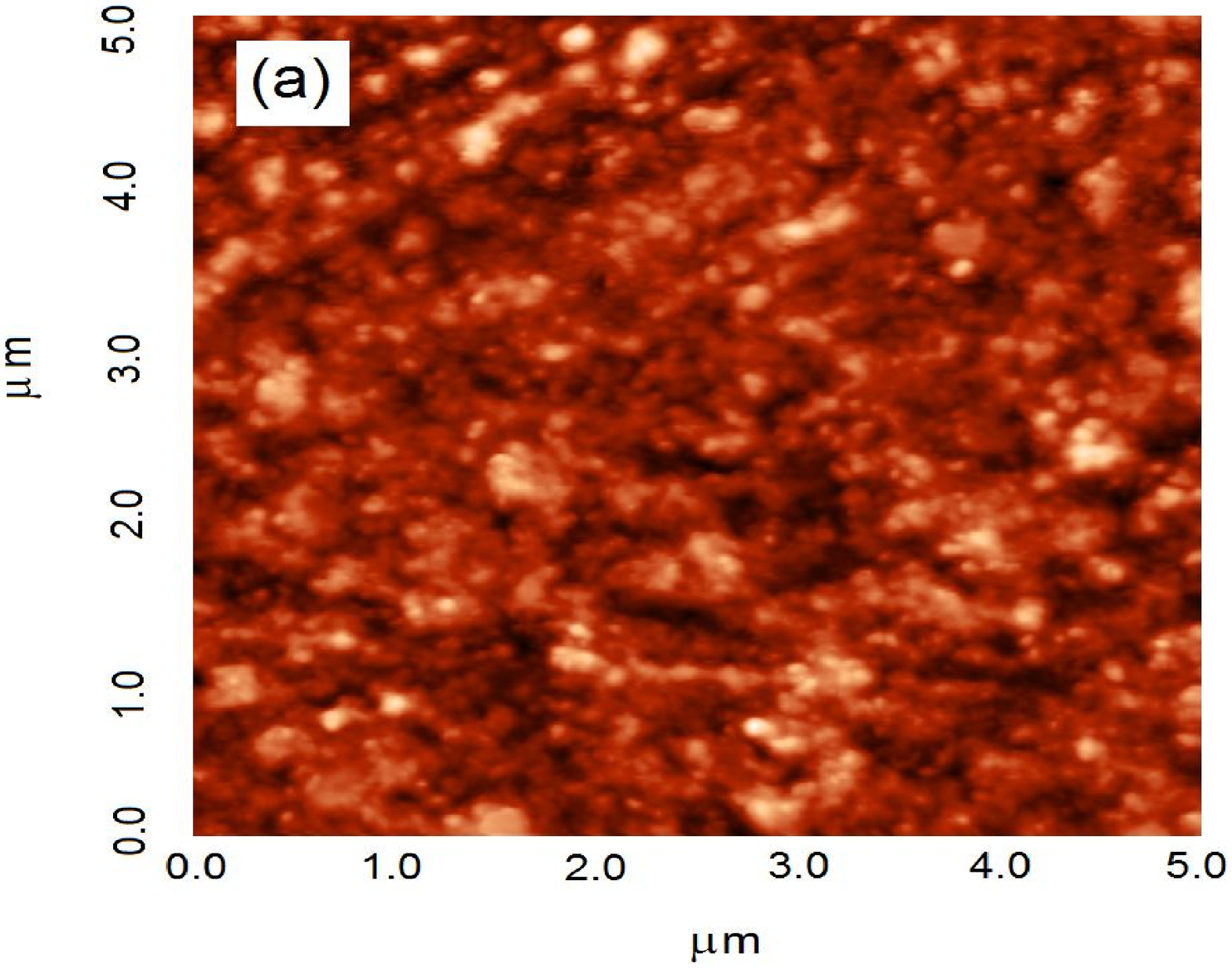, width=3in, angle=0}
\vfil
\epsfig{file=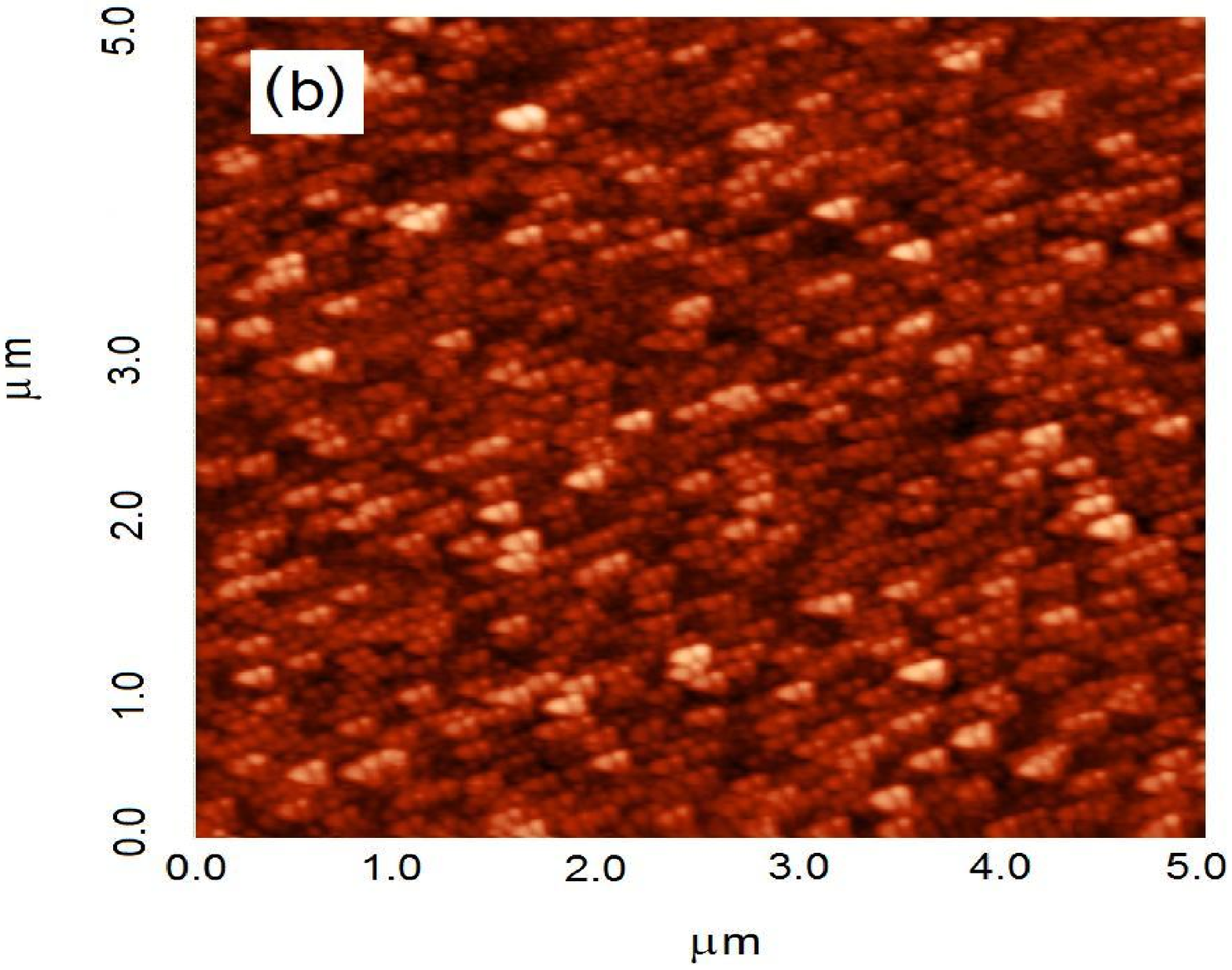, width=3in, angle=0}
\end{center}
\caption{\sl AFM images of sample (a) and sample (b).}
\end{figure}

\begin{figure}[h]
\begin{center}
\epsfig{file=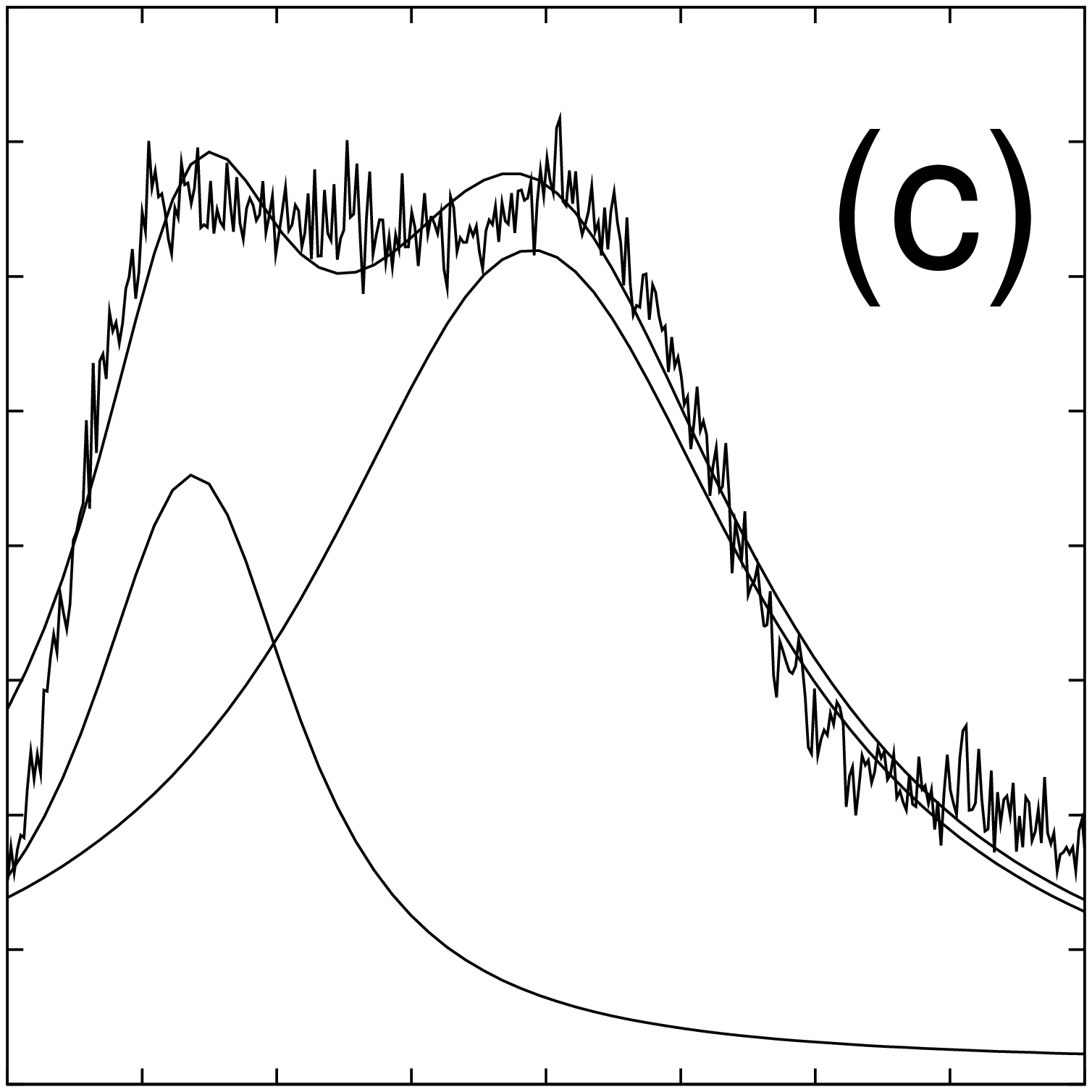, width=1.15in, angle=-90}
\vfil
\epsfig{file=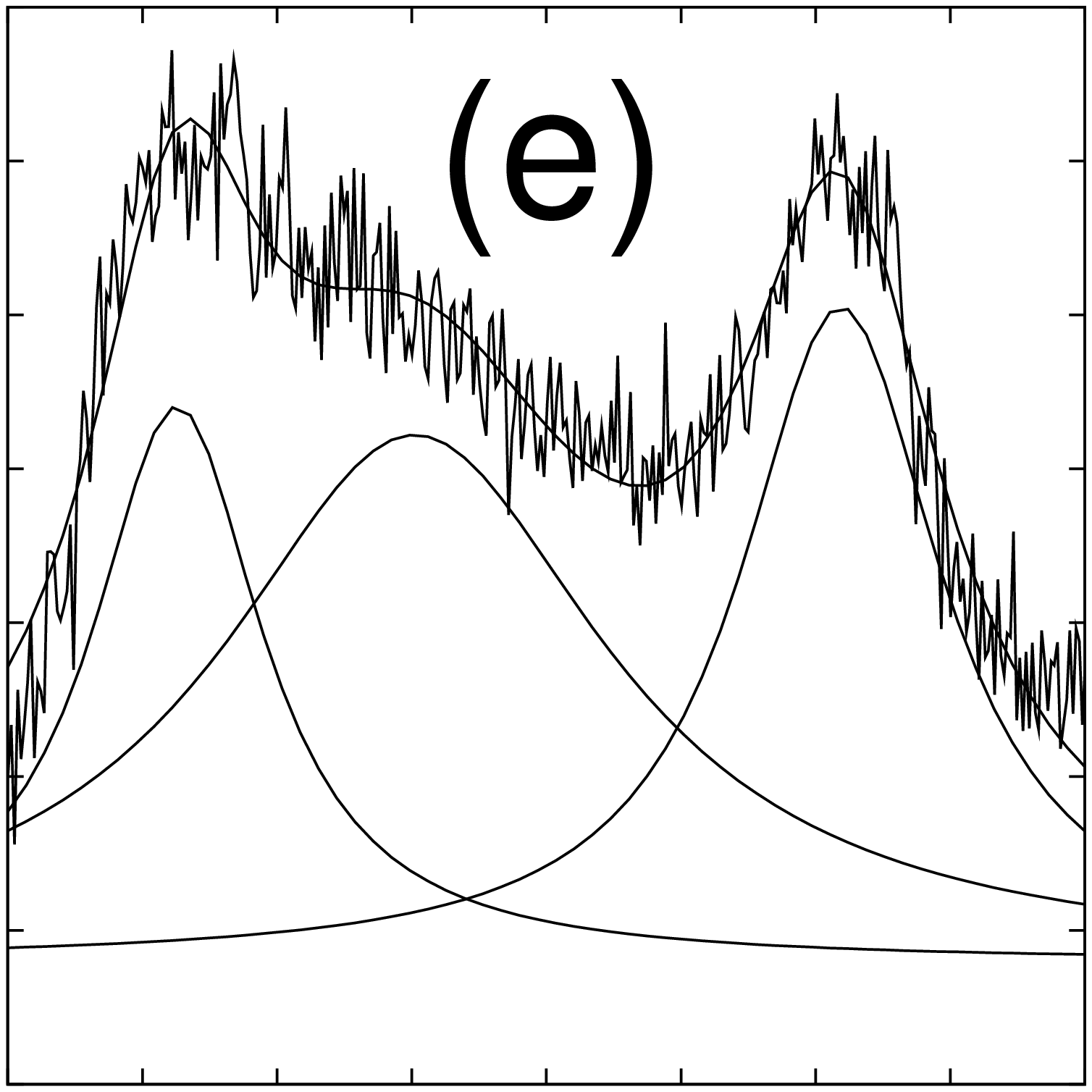, width=1.15in, angle=-90}
\vfil
\epsfig{file=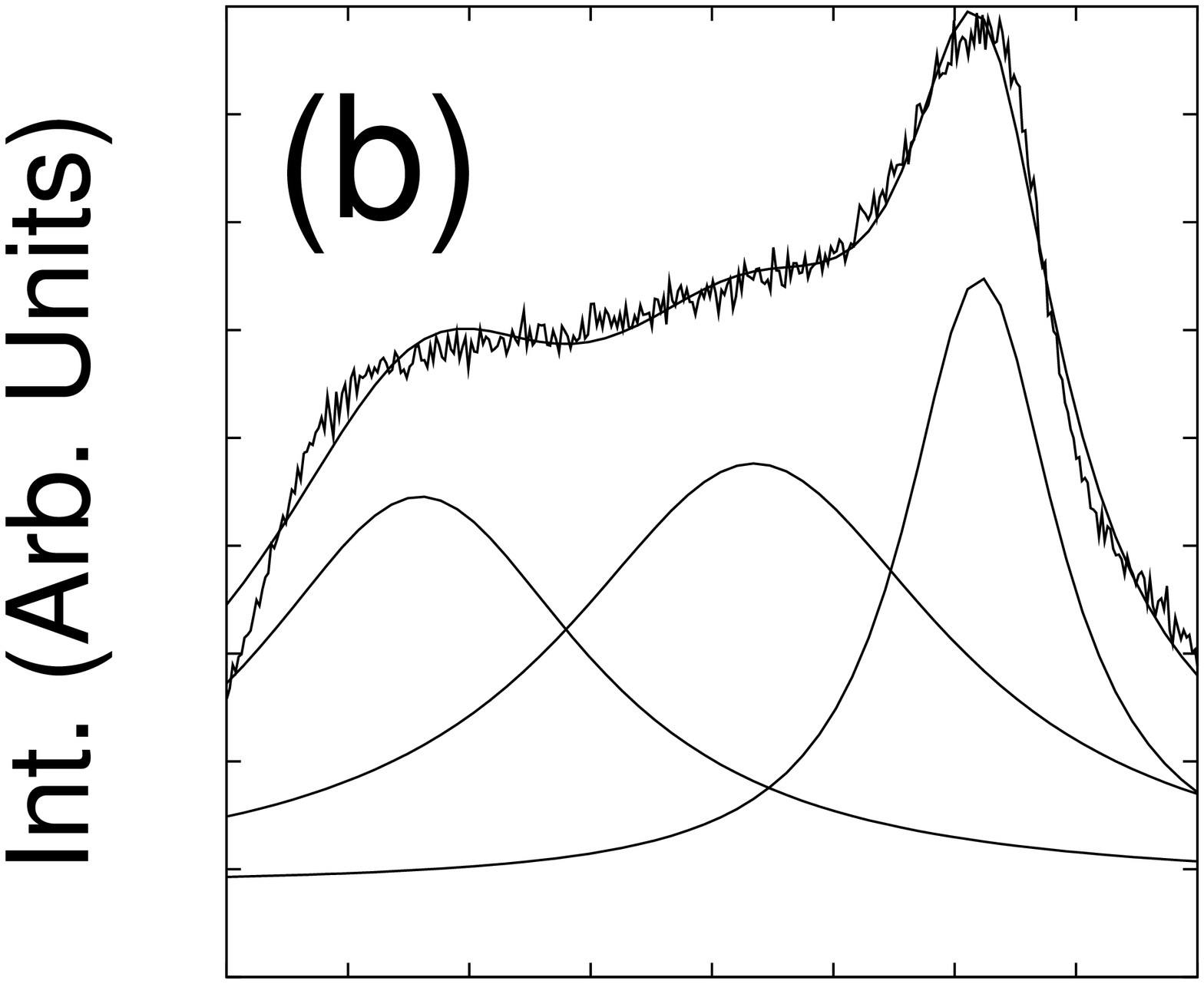, width=1.15in, angle=-90}
\vfil
\epsfig{file=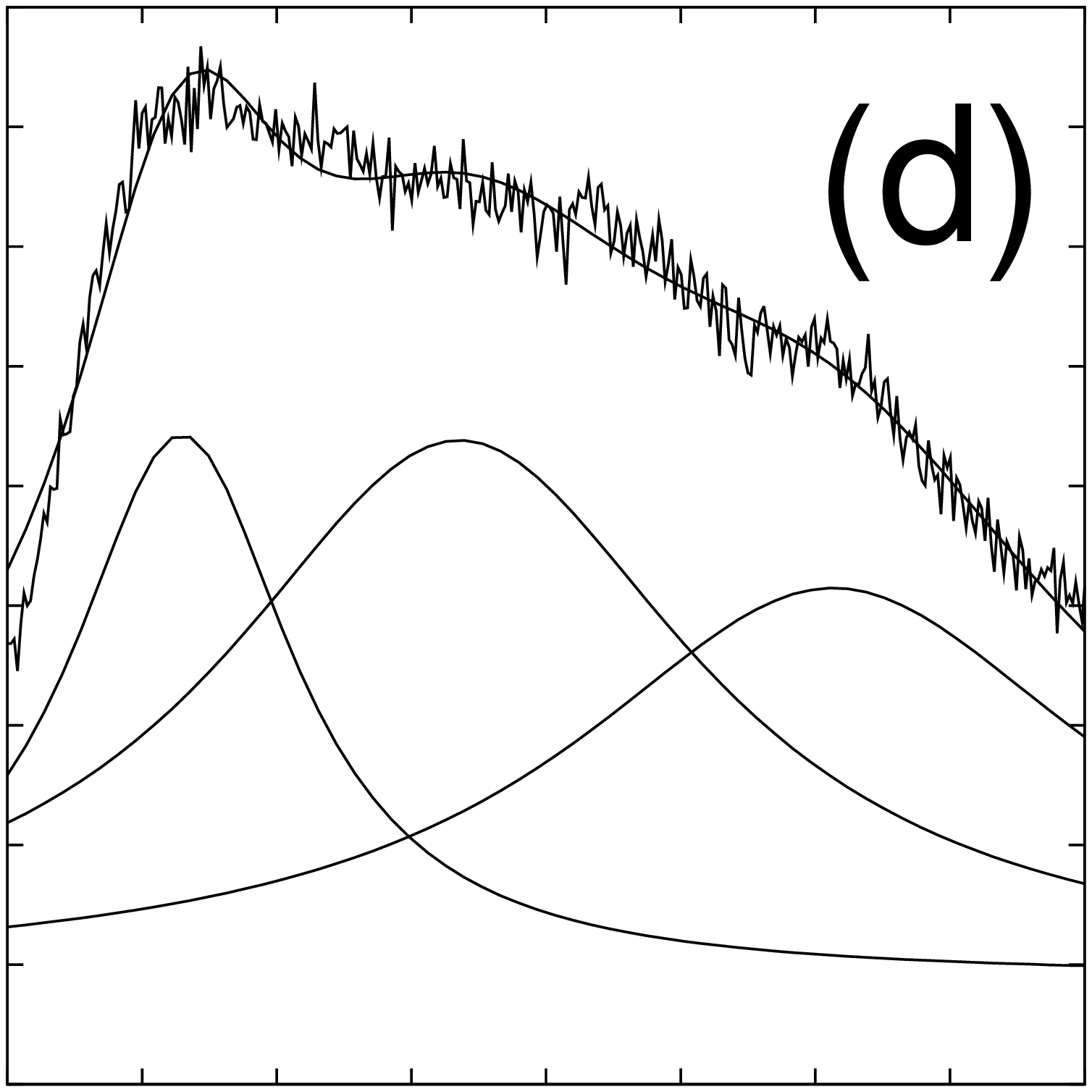, width=1.15in, angle=-90}
\vfil
\epsfig{file=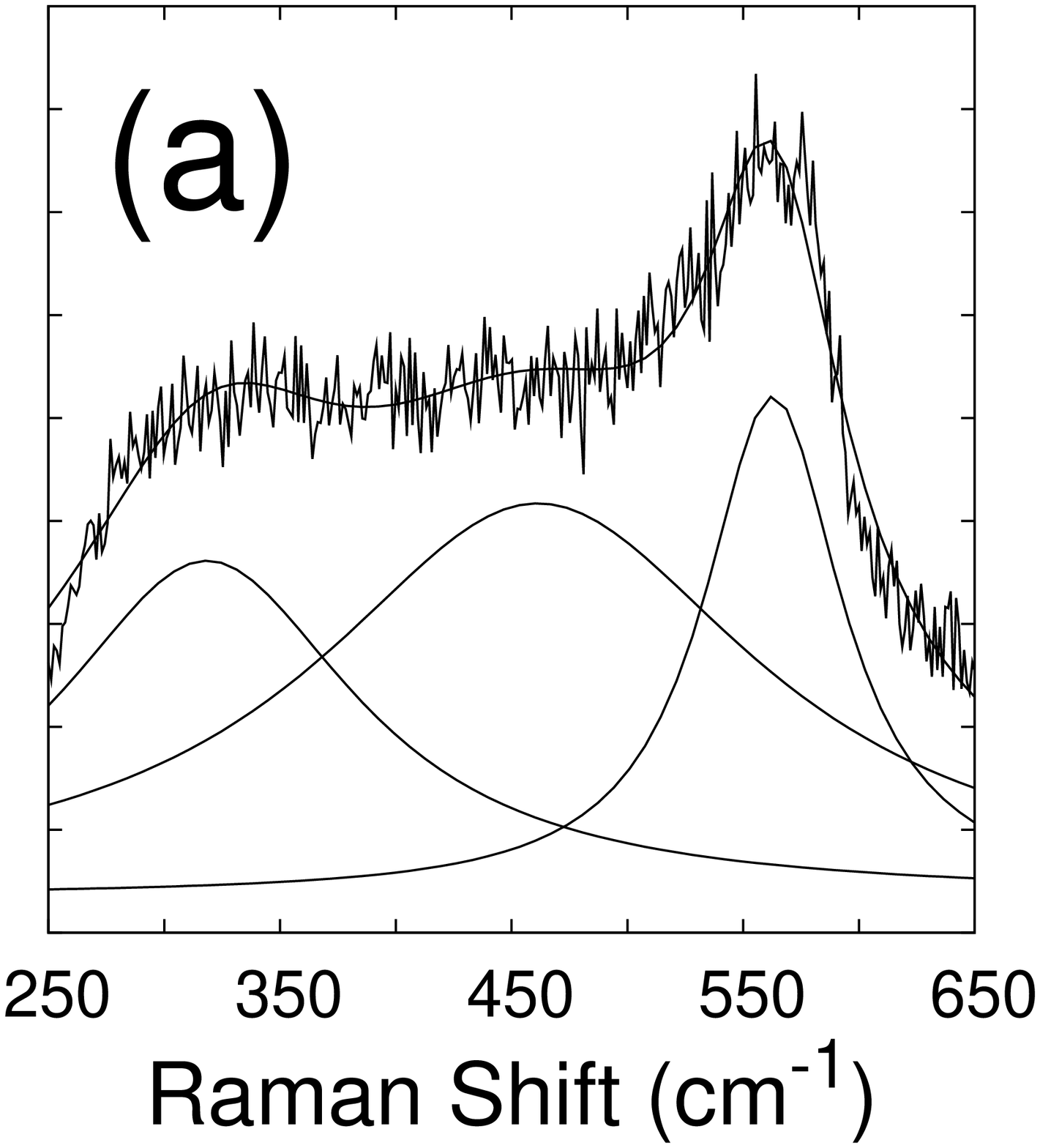, width=1.15in, angle=-90}
\end{center}
\caption{\sl Raman spectra of sample (a), (d), (b), (e) and (c). Also seen 
are deconvoluted peaks assigned to amorphous silicon, wurtzite structure ZnO 
and with oxygen vacancies defects.}
\end{figure}

\begin{figure}[h]
\begin{center}
\epsfig{file=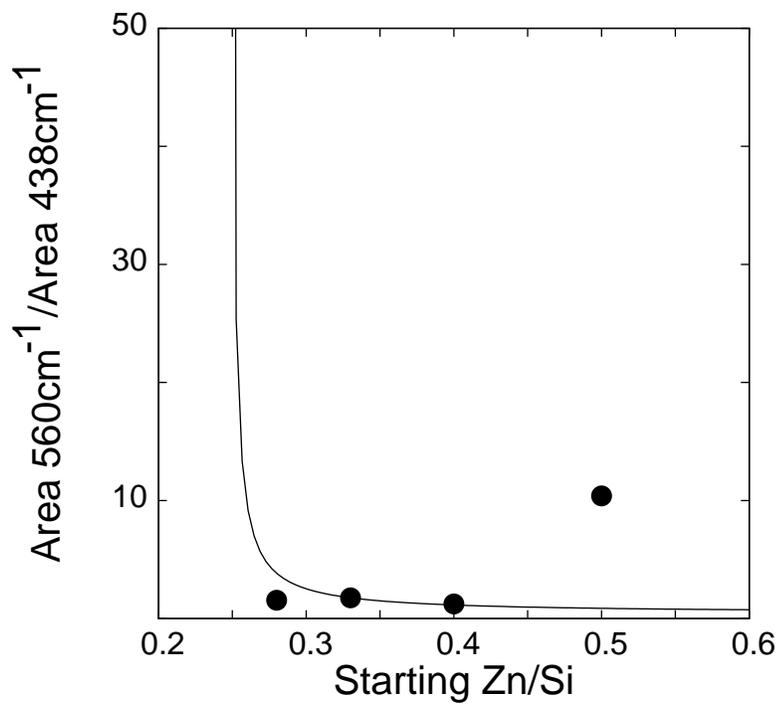, width=3.5in, angle=-90}
\end{center}
\caption{\sl 
Relative presence of ZnO with oxygen vacancies to wurtzite 
structure ZnO (Area ${\rm 560cm^{-1}}$/Area ${\rm 438cm^{-1}}$ from Raman 
spectra) for varying ZnO content in film.
}
\end{figure}

\begin{figure}[h]
\begin{center}
\epsfig{file=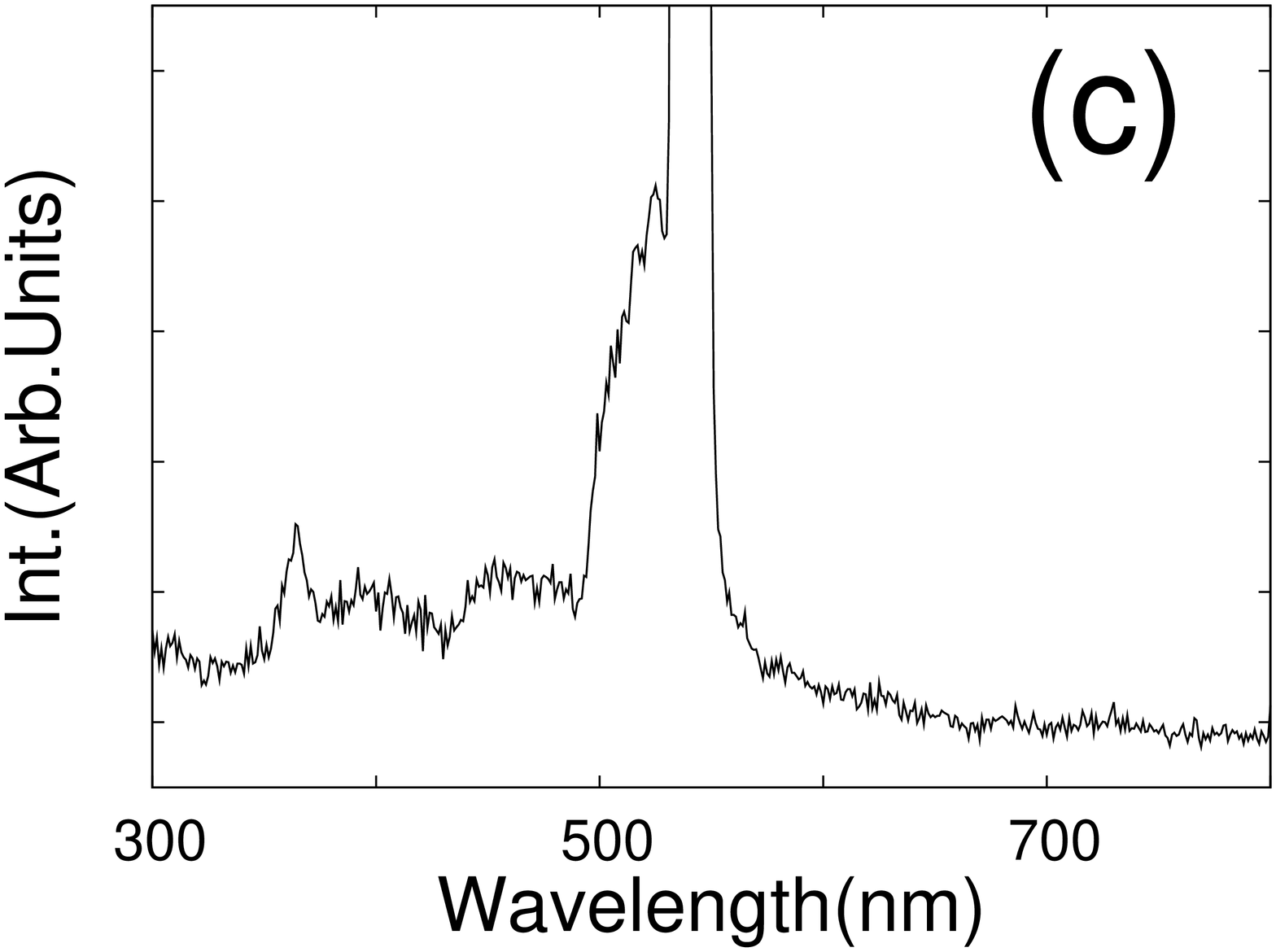, width=1.5in, angle=-90}
\hfil
\epsfig{file=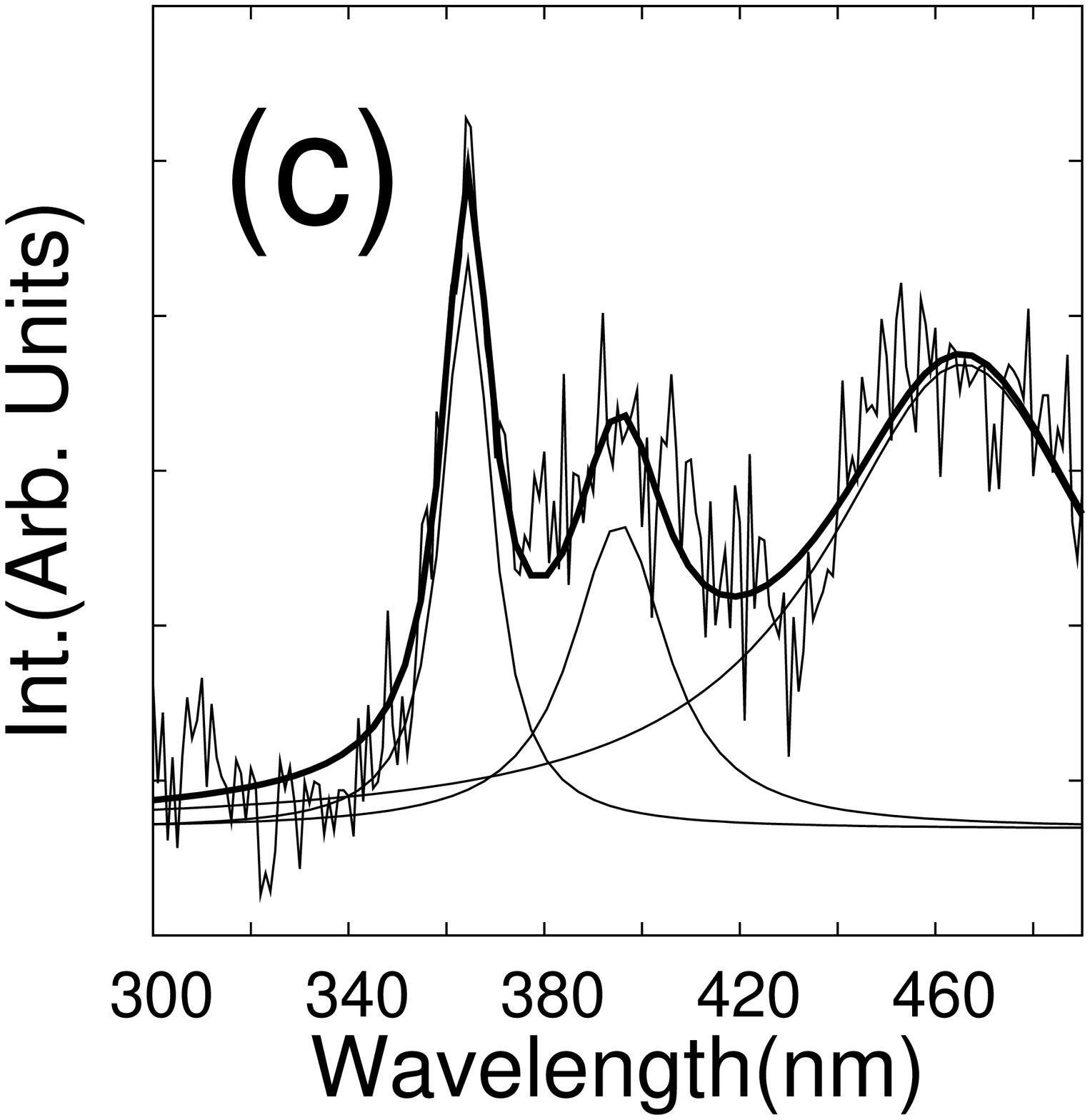, width=1.5in, angle=-90}
\hfil
\epsfig{file=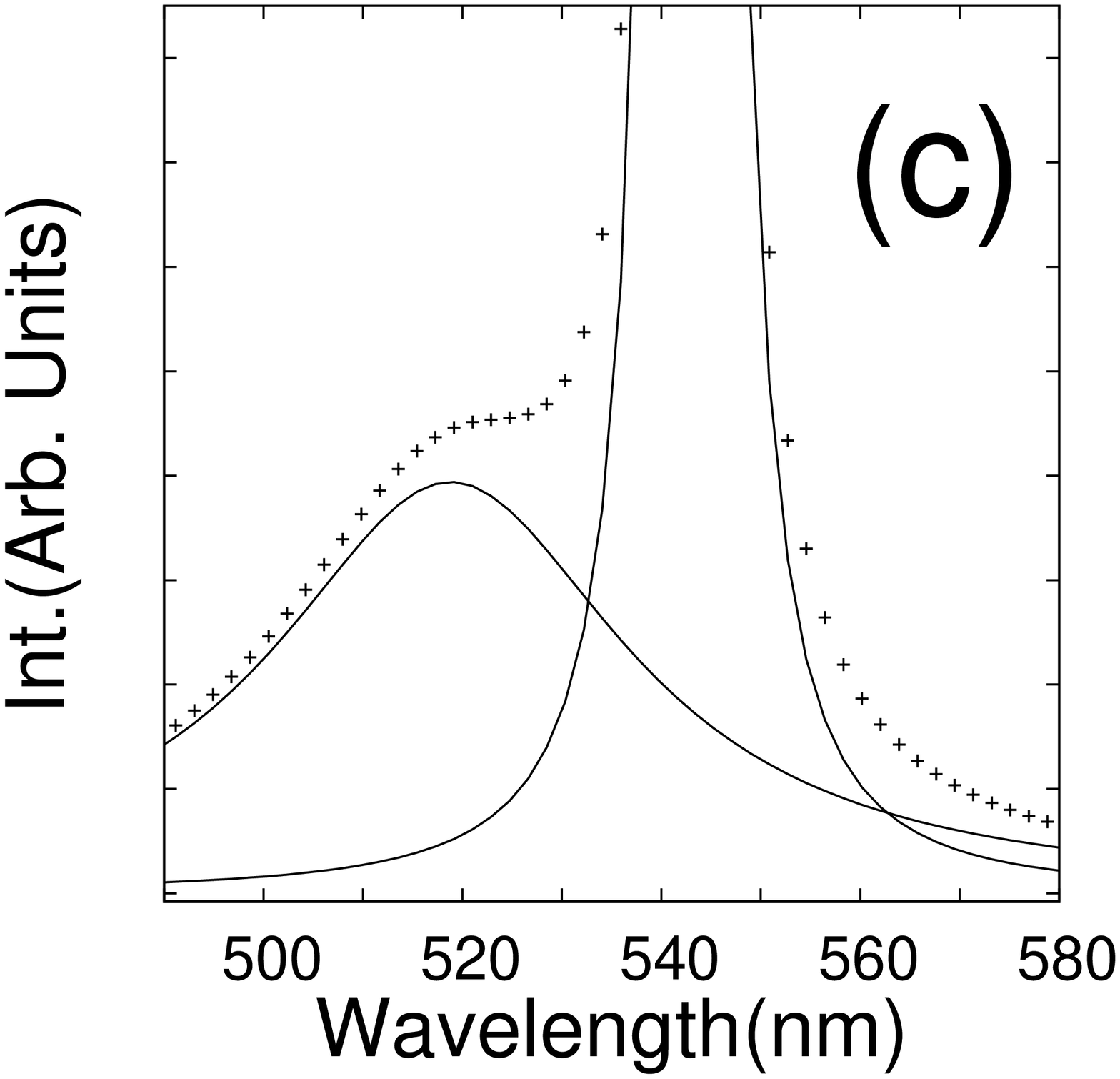, width=1.5in, angle=-90}
\vfil
\vskip 0.5cm
\epsfig{file=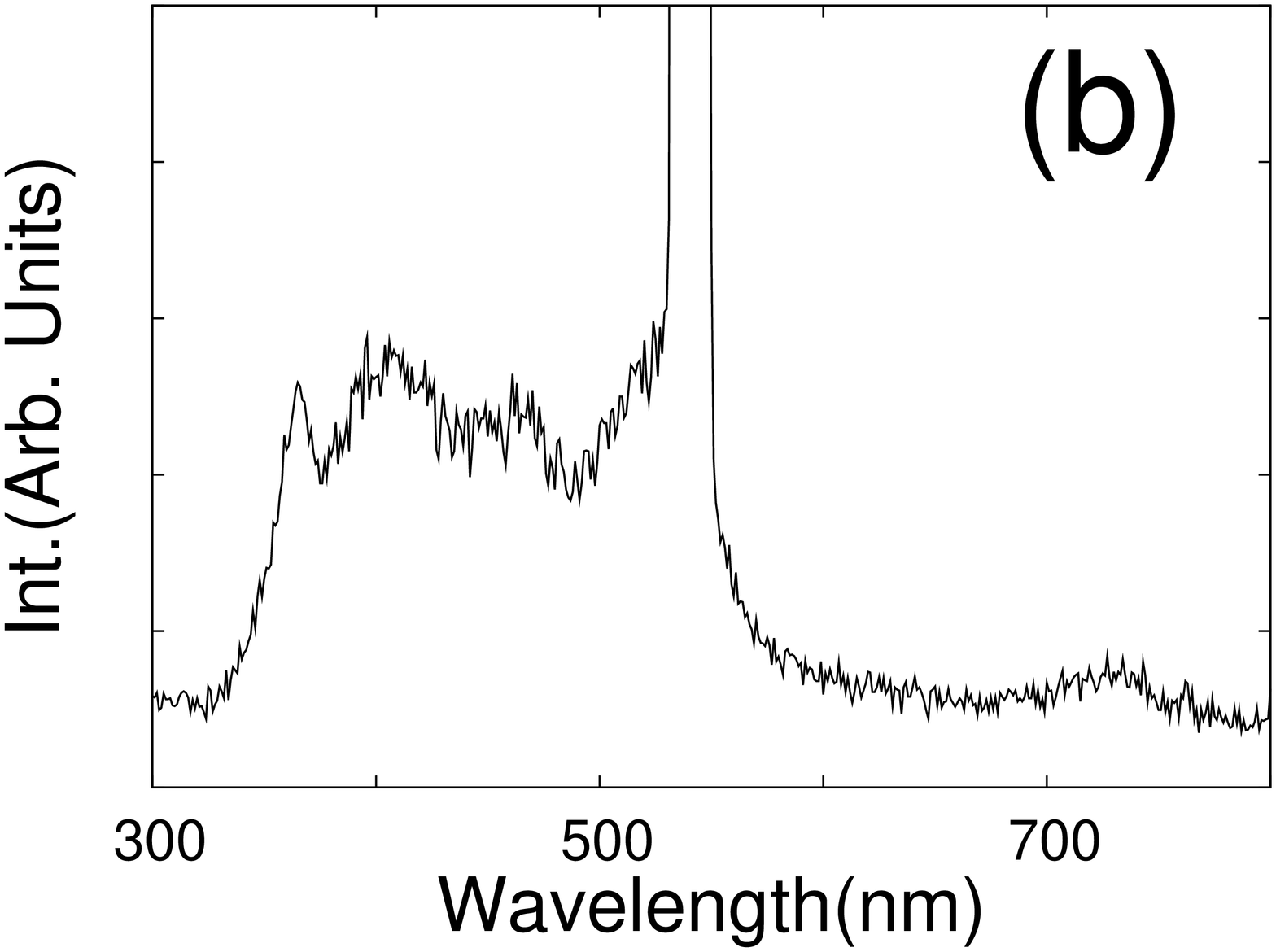, width=1.5in, angle=-90}
\hfil
\epsfig{file=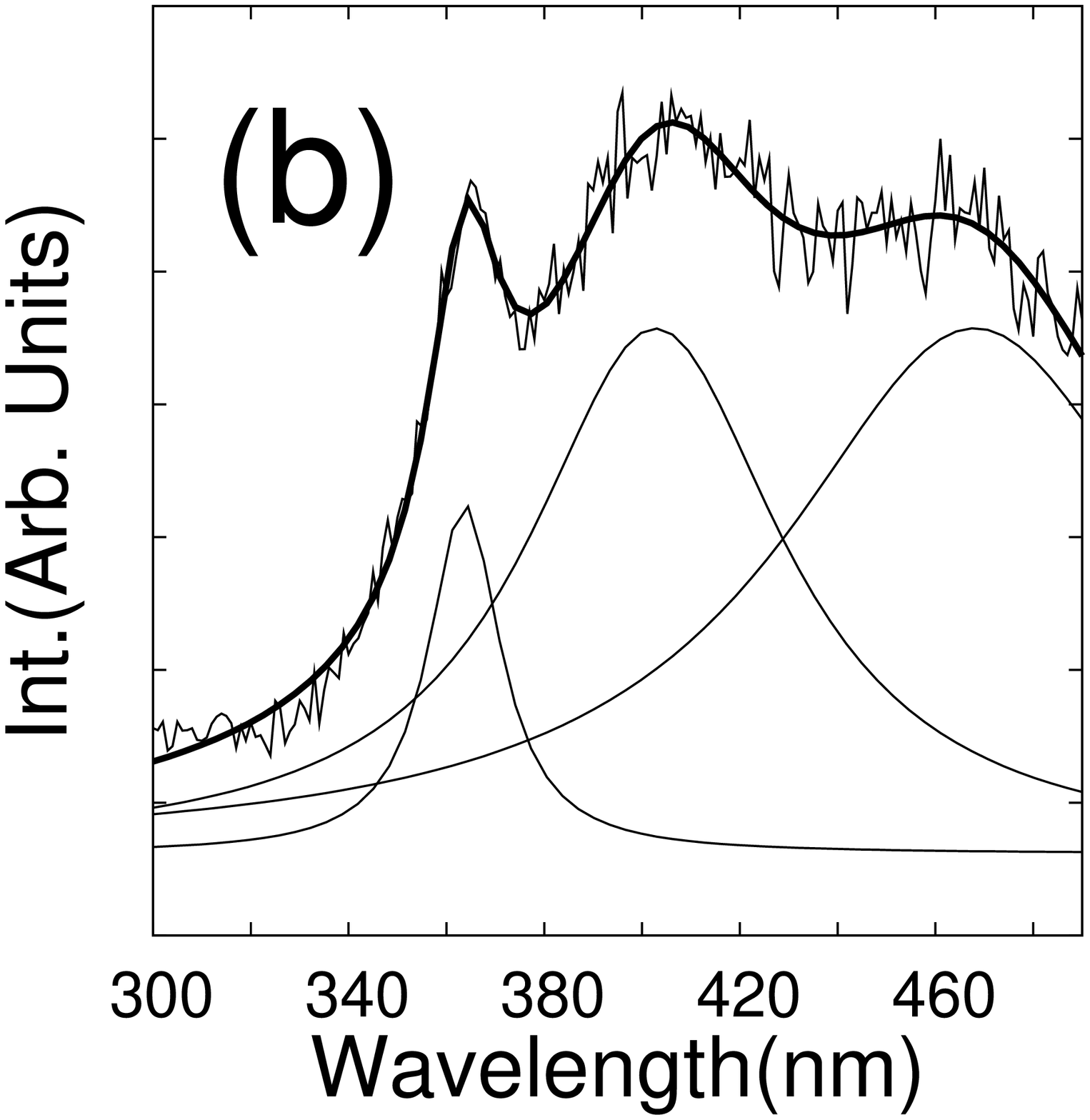, width=1.5in, angle=-90}
\hfil
\epsfig{file=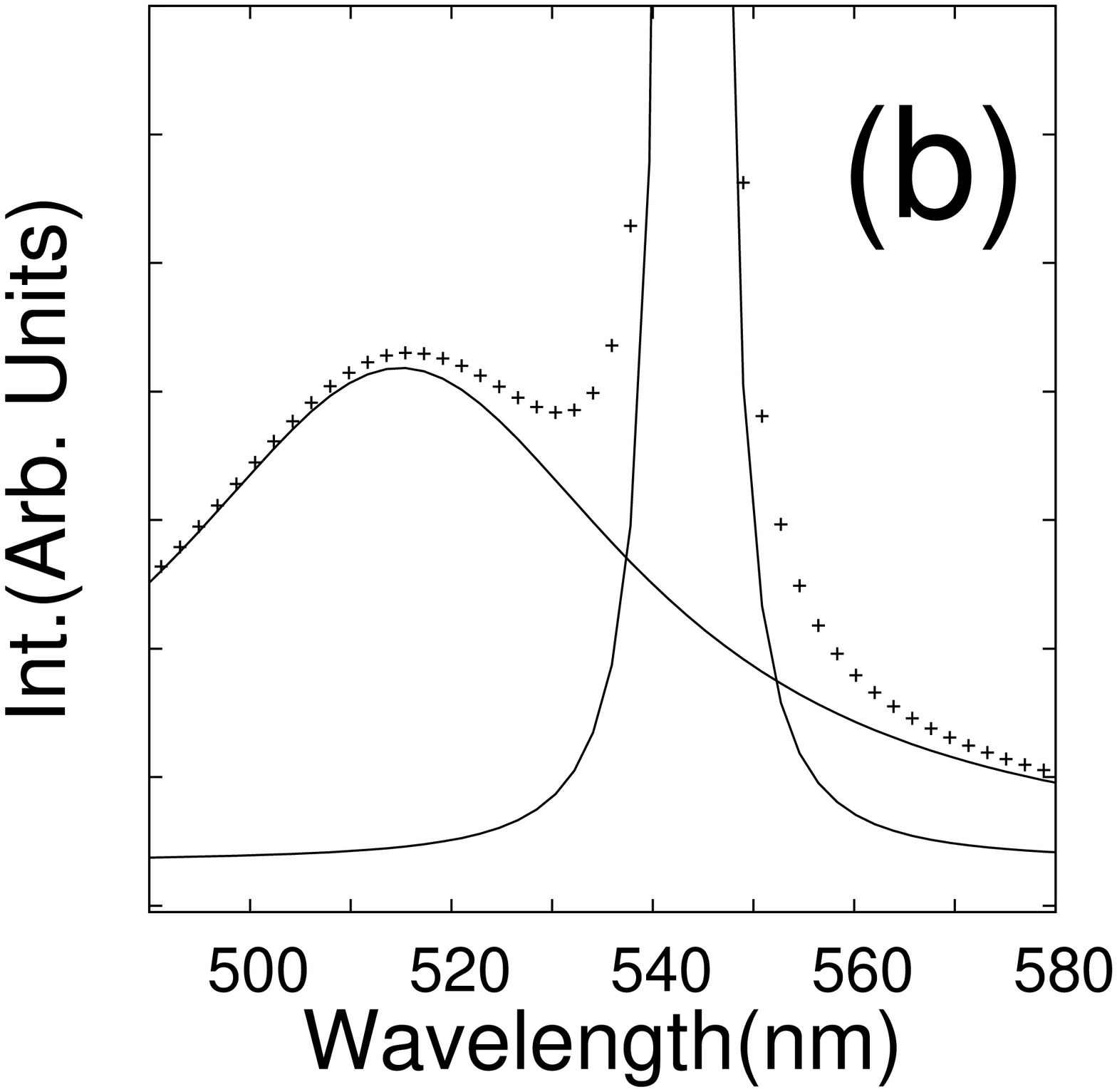, width=1.5in, angle=-90}
\vfil
\vskip 0,5cm
\epsfig{file=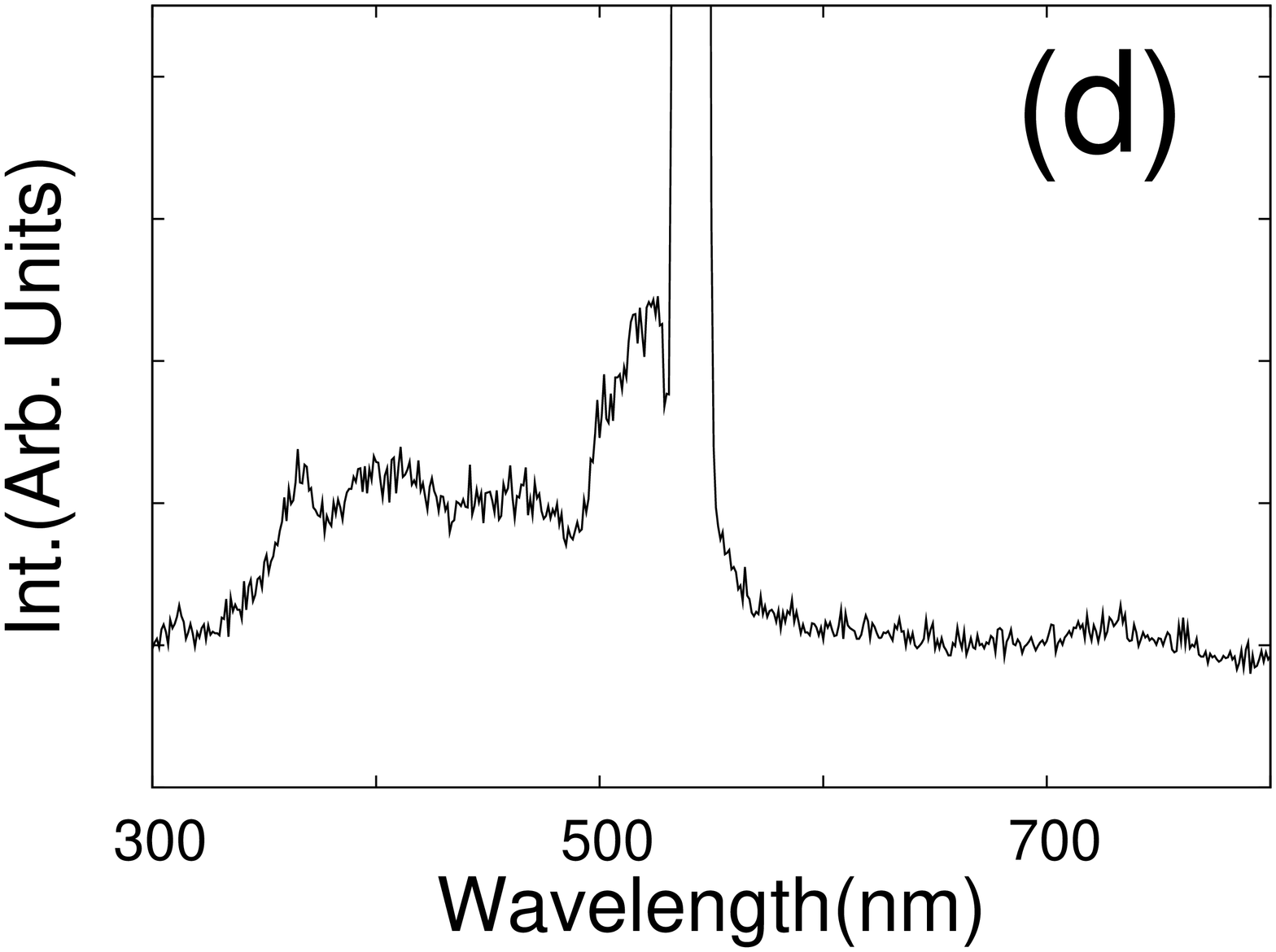, width=1.5in, angle=-90}
\hfil
\epsfig{file=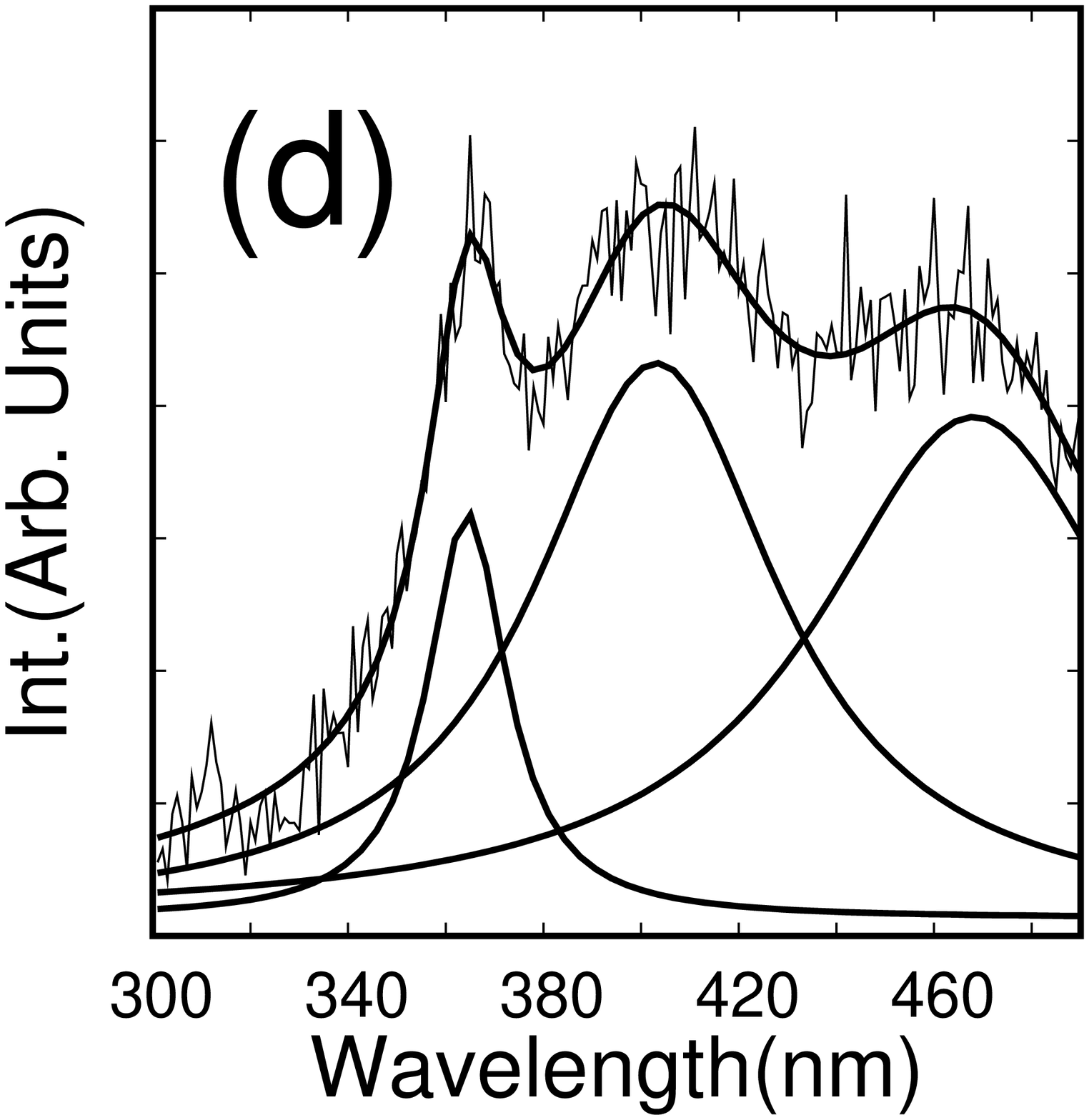, width=1.5in, angle=-90}
\hfil
\epsfig{file=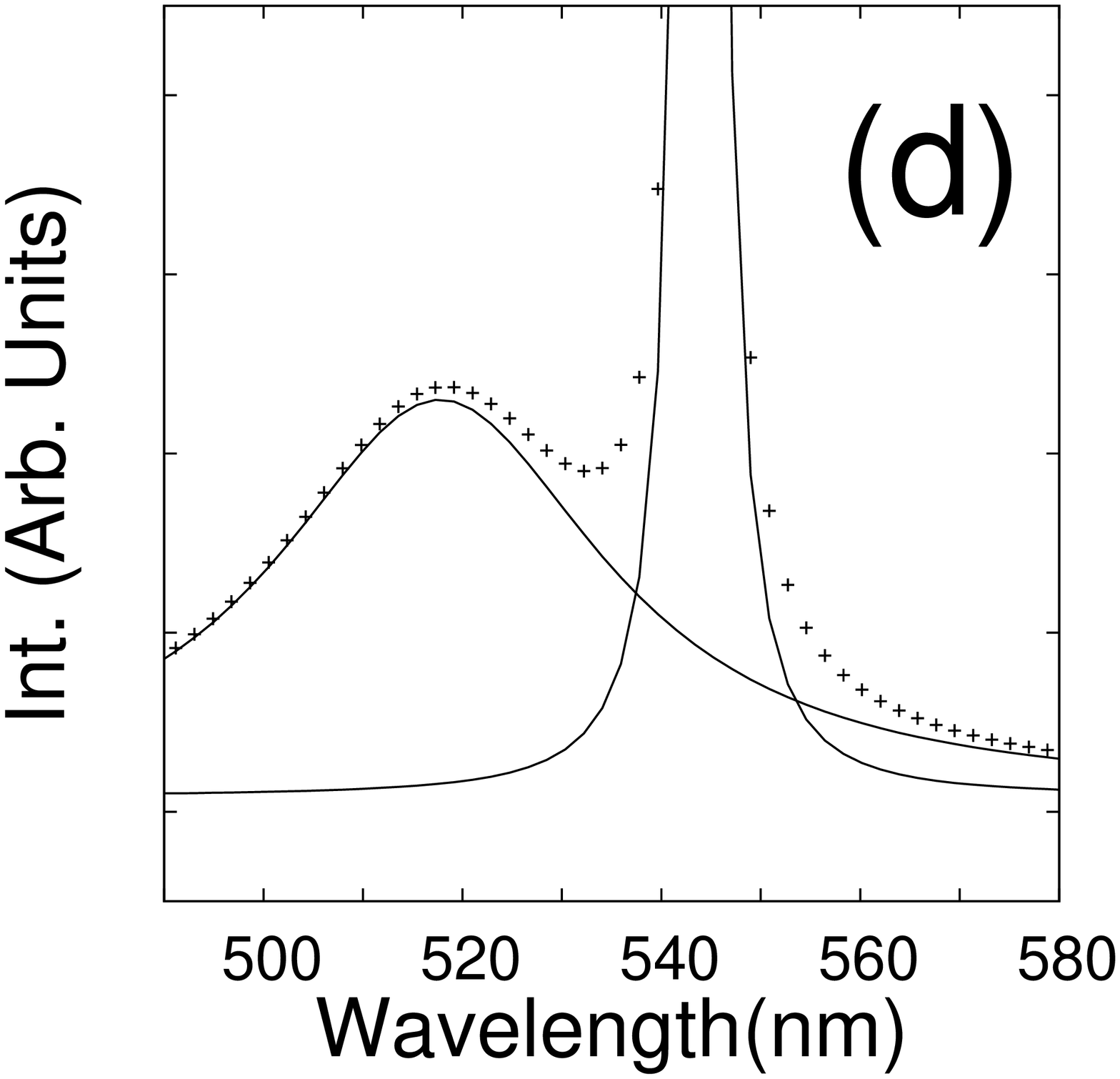, width=1.5in, angle=-90}
\vfil
\vskip 0.5cm
\epsfig{file=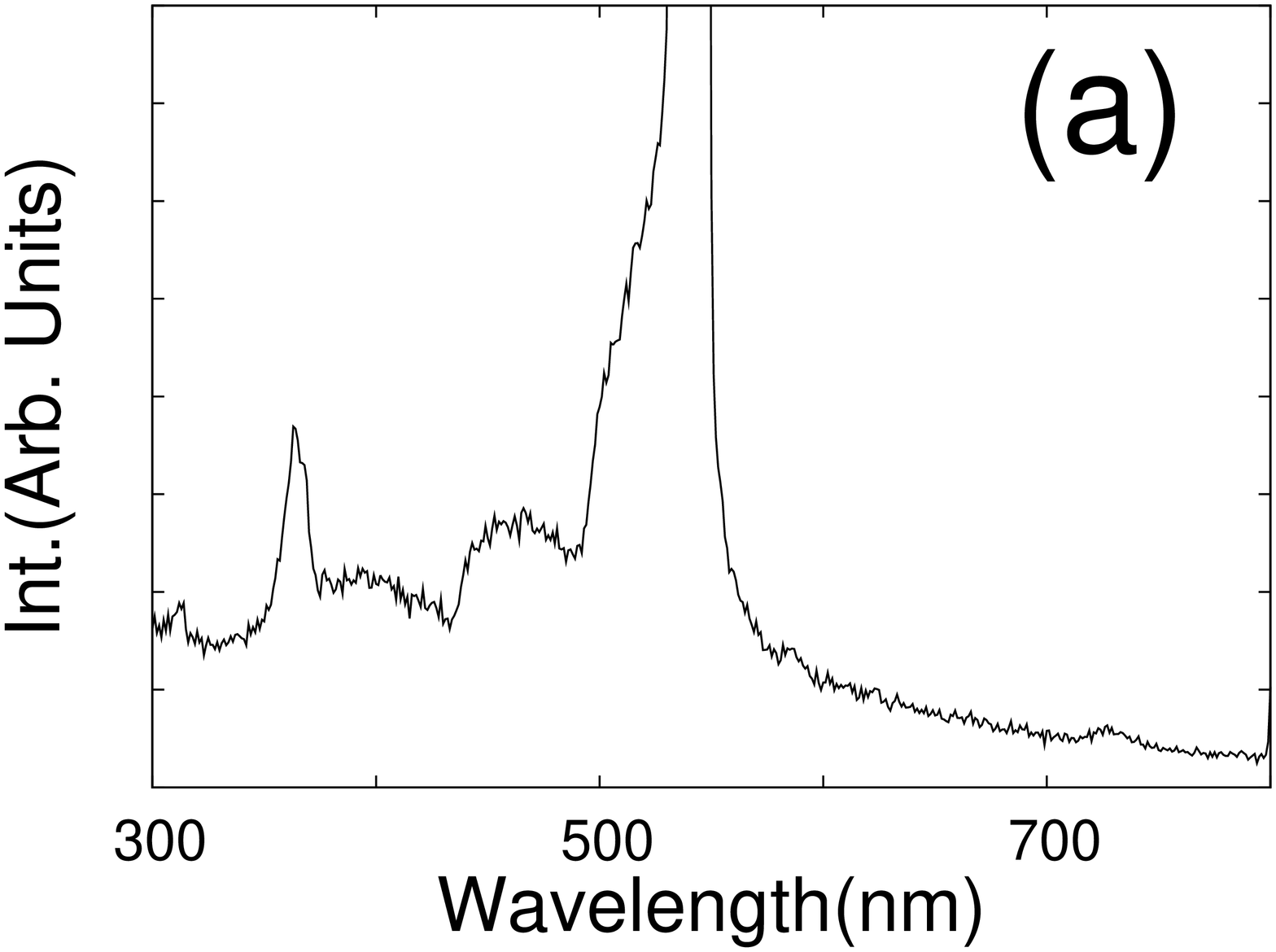, width=1.5in, angle=-90}
\hfil
\epsfig{file=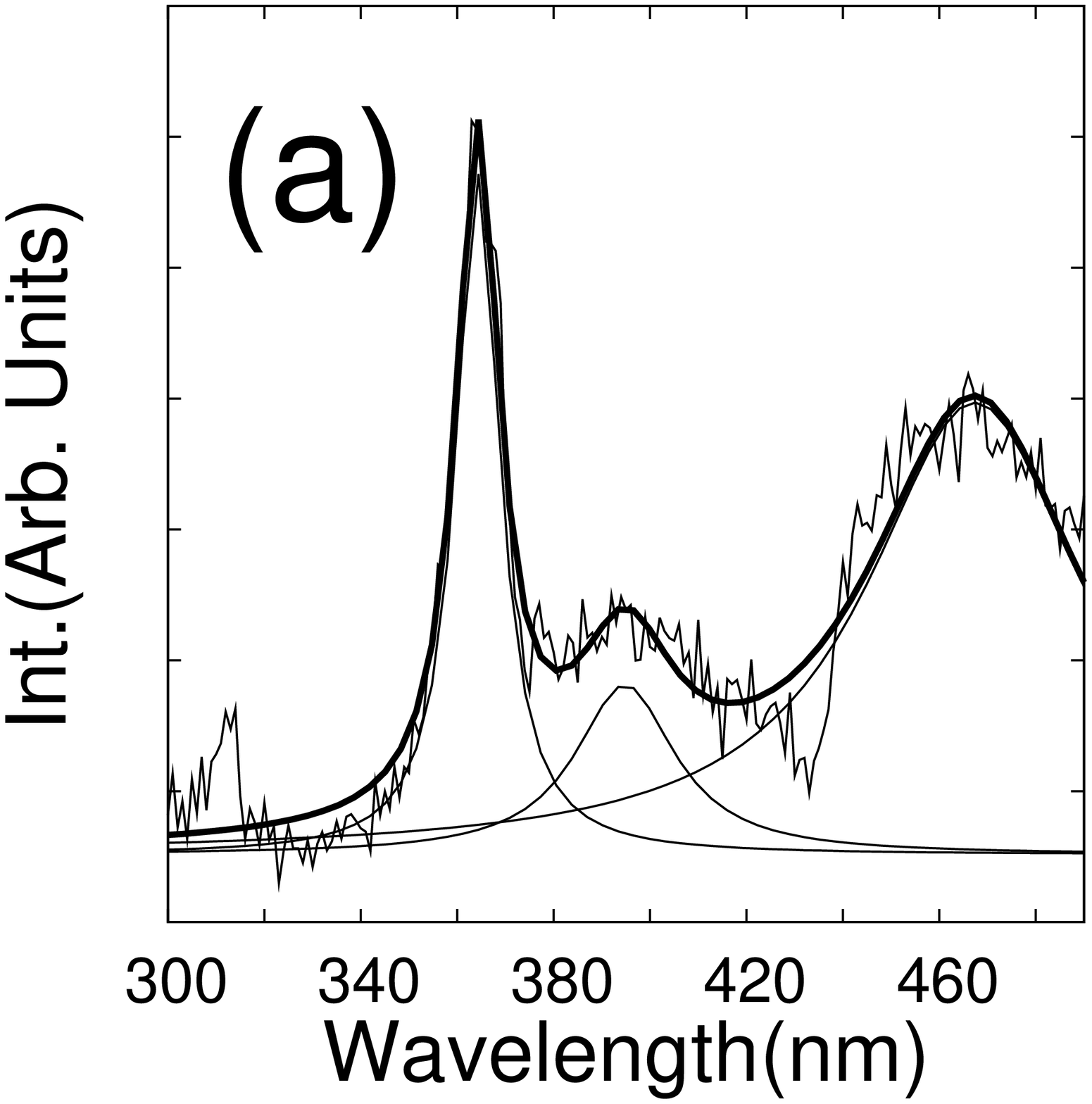, width=1.5in, angle=-90}
\hfil
\epsfig{file=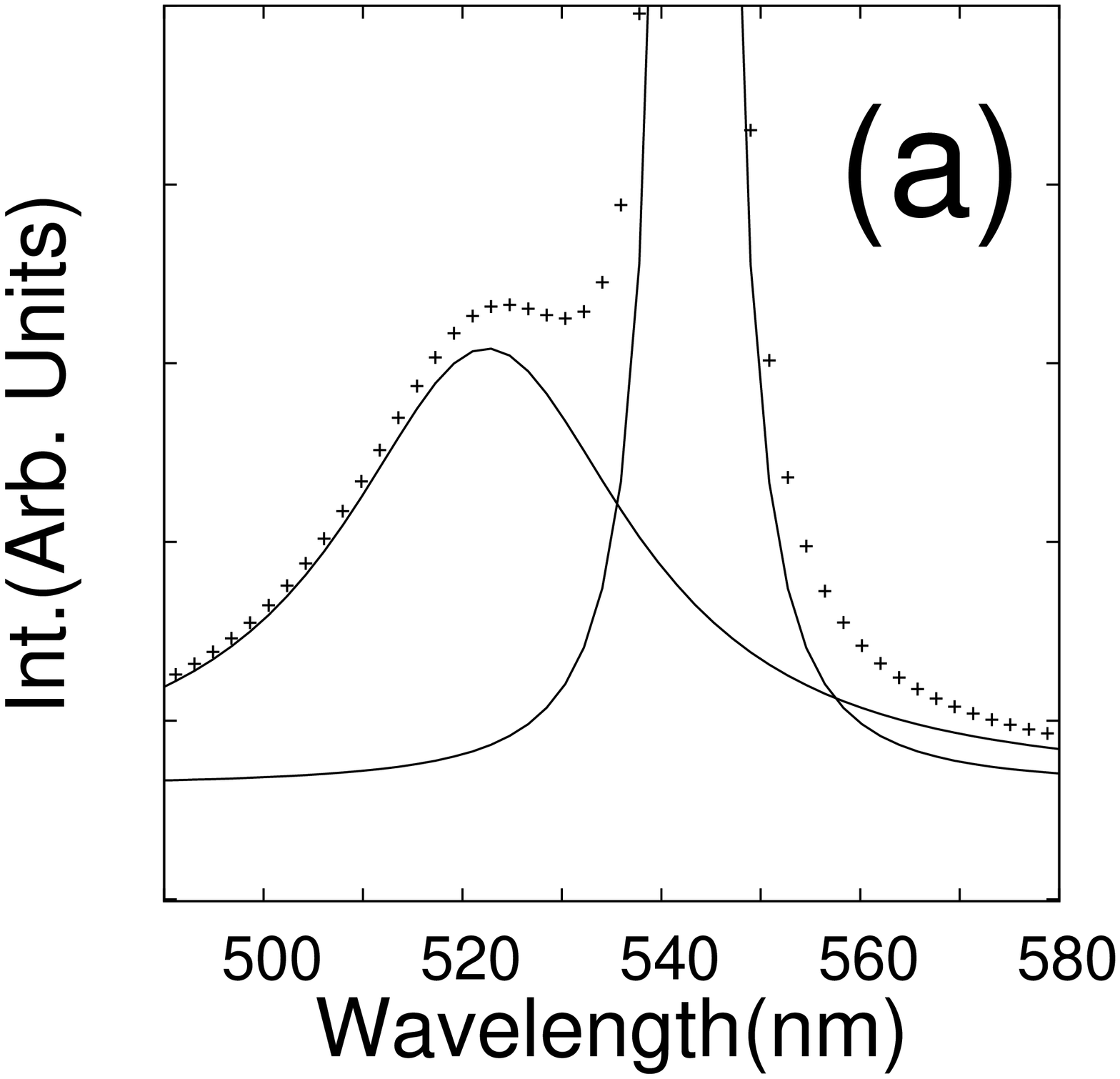, width=1.5in, angle=-90}
\end{center}
\caption{\sl PL of sample (a), (d), (b), and (c). 
Alongside the raw spectra are shown, deconvolution give 365, 400 and 
465nm between 300 and 480nm. Also green emission due to defects have 
been separated from $\rm 2^{nd}$ harmonic to show relative contributions.}
\end{figure}

\begin{figure}[h]
\begin{center}
\epsfig{file=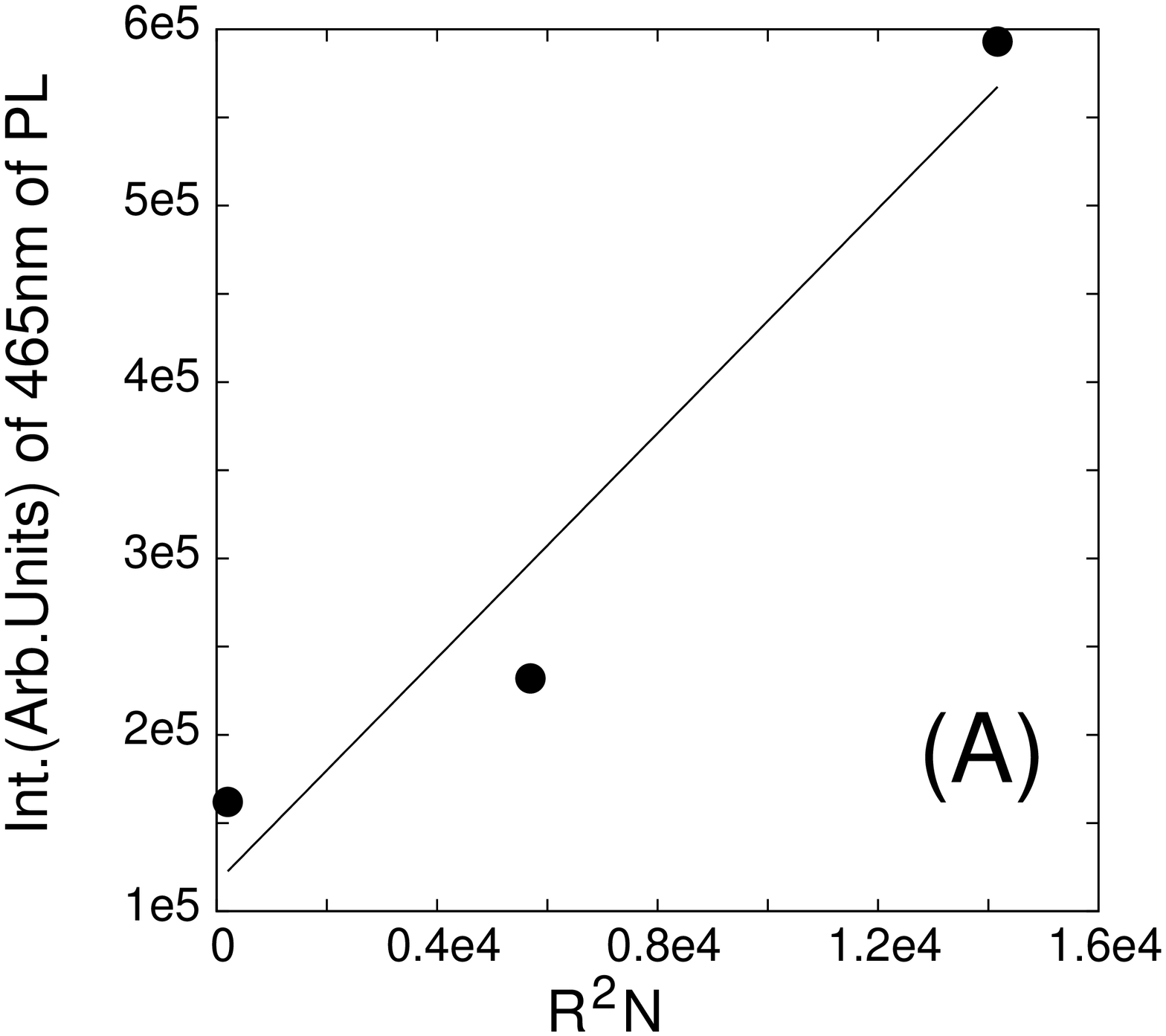, width=2.05in, angle=-90}
\vfil
\vskip 0.7cm
\epsfig{file=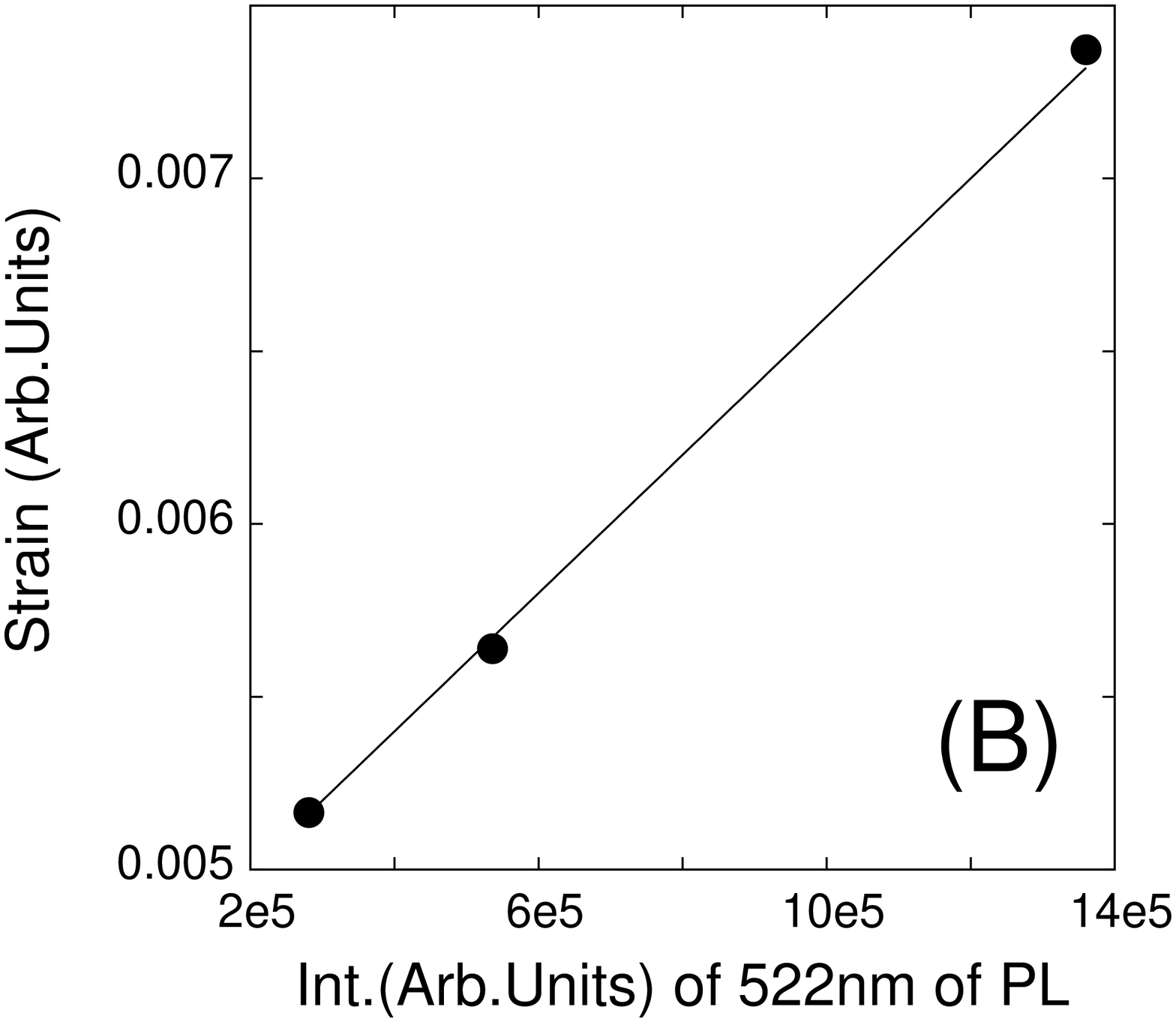, width=2.05in, angle=-90}
\vfil
\vskip 0.7cm
\epsfig{file=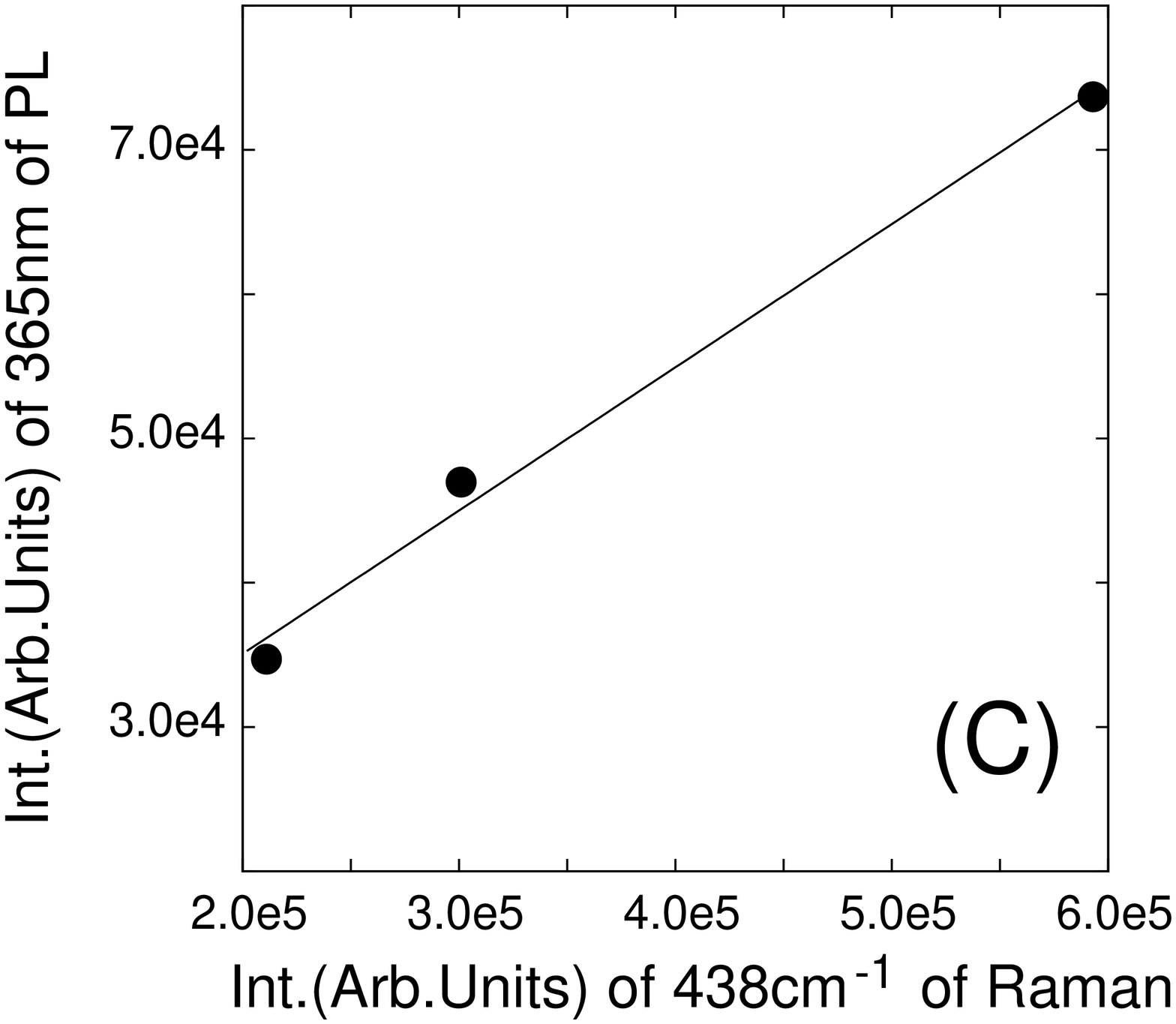, width=2.05in, angle=-90}
\end{center}
\caption{\sl The (A) increase in contribution of 465nm PL peak with 
increasing ${\rm R^2N}$, (B) linear relation in green (522nm) emission with 
strain in film and inturn oxygen vacancies and (C) co-relation in existence 
of wurtzite peak (${\rm 438cm^{-1}}$ in Raman spectra) and blue emision 
(365nm peak of PL) in samples.}
\end{figure}

\begin{figure}[h]
\begin{center}
\epsfig{file=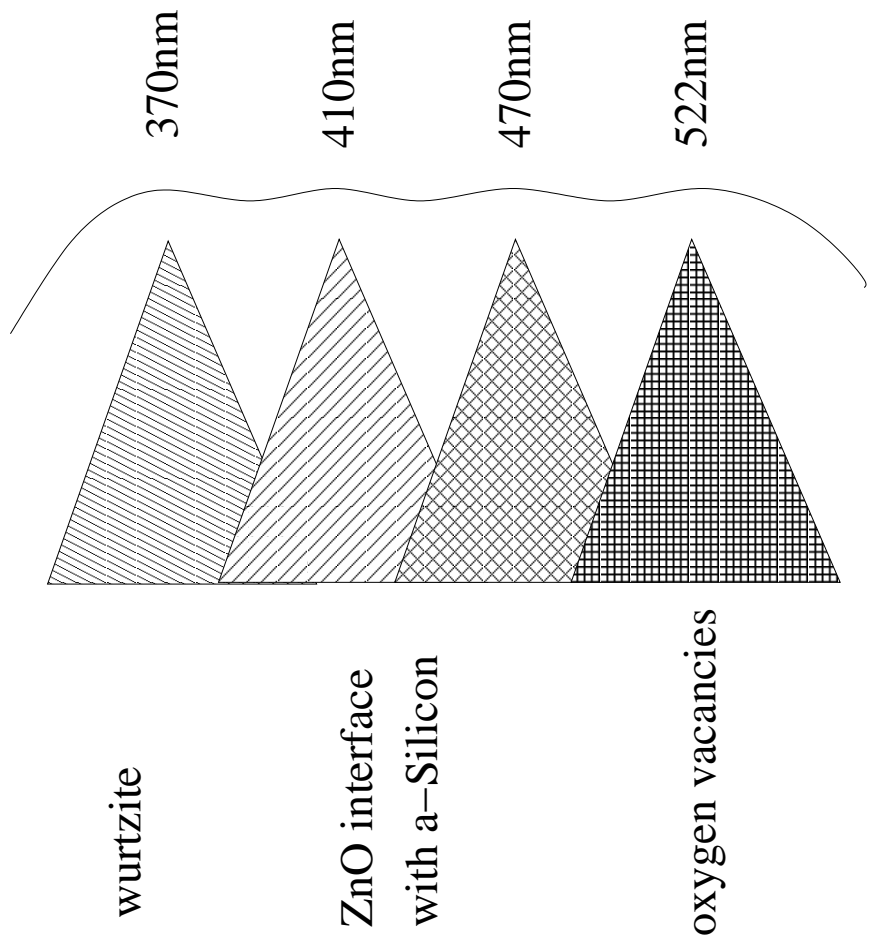, width=4.5in, angle=-0}
\end{center}
\caption{\sl {Schematic explaining broadening of PL in ZnO:Si nanocomposites 
and their individual contributions.}.}
\end{figure}

\end{document}